\documentclass[prb,a4paper,twocolumn,showpacs,floatfix,superscriptaddress,amsmath,amsfonts,amssymb,preprintnumbers,citeautoscript]{revtex4-2}
\pdfoutput=1 \widowpenalty10000 \clubpenalty10000
\usepackage[T1]{fontenc}\usepackage[latin1]{inputenc}
\usepackage{dcolumn,graphicx,color,booktabs,microtype,afterpage} \graphicspath{{Figures/}}
\usepackage[charter,greekuppercase=italicized]{mathdesign}\usepackage{sidecap}

\newcount\hh \newcount\mm
\hh=\time \divide\hh by 60
\mm=\hh \multiply\mm by 60 \mm=-\mm
\advance\mm by \time
\def\now{\number\hh:\ifnum\mm<10{}0\fi\number\mm}

\usepackage[colorlinks,plainpages=false,linkcolor=blue,urlcolor=blue,citecolor=blue,pdfpagemode=UseNone,pdfstartview=FitBH]{hyperref}

\newcommand{\jt}{\ensuremath{J_{\text{t}}}}
\newcommand{\jc}{\ensuremath{J_{\text{c}}}}
\newcommand{\jtt}{\ensuremath{J_{\text{tt}}}}
\newcommand{\jct}{\ensuremath{J_{\text{ct}}}}
\newcommand{\jp}{\ensuremath{J_{\perp}}}

\begin{document}

\makeatletter\renewcommand{\ps@plain}{%
	\def\@evenhead{\hfill\itshape\rightmark}%
	\def\@oddhead{\itshape\leftmark\hfill}%
	\renewcommand{\@evenfoot}{\hfill\small{--~\thepage~--}\hfill}%
	\renewcommand{\@oddfoot}{\hfill\small{--~\thepage~--}\hfill}%
}\makeatother\pagestyle{plain}


\title{Destruction of long-range magnetic order in an external magnetic field\\and the associated spin dynamics in Cu$_2$GaBO$_5$ and Cu$_2$AlBO$_5$ ludwigites.}

\author{A.~A.~Kulbakov}
\affiliation{Institut f\"ur Festk\"orper- und Materialphysik, Technische Universit\"at Dresden, 01069 Dresden, Germany}
\affiliation{W\"urzburg-Dresden Cluster of Excellence on Complexity and Topology in Quantum Matter\,---\,\textit{ct.qmat}, TU Dresden, 01069 Dresden, Germany}

\author{R.~Sarkar}
\affiliation{Institut f\"ur Festk\"orper- und Materialphysik, Technische Universit\"at Dresden, 01069 Dresden, Germany}

\author{O.~Janson}
\affiliation{Institute for Theoretical Solid State Physics, IFW Dresden, 01069 Dresden, Germany}

\author{S.~Dengre}
\affiliation{Institut f\"ur Festk\"orper- und Materialphysik, Technische Universit\"at Dresden, 01069 Dresden, Germany}

\author{T.~Weinhold}
\affiliation{Institut f\"ur Festk\"orper- und Materialphysik, Technische Universit\"at Dresden, 01069 Dresden, Germany}

\author{E.~M.~Moshkina}
\affiliation{Kirensky Institute of Physics, Federal Research Center KSC SB RAS, Akademgorodok~50, 660036 Krasnoyarsk, Russia}

\author{P.~Y.~Portnichenko}
\affiliation{Institut f\"ur Festk\"orper- und Materialphysik, Technische Universit\"at Dresden, 01069 Dresden, Germany}

\author{H.~Luetkens}
\affiliation{Laboratory for Muon Spin Spectroscopy, Paul Scherrer Institute, CH-5232 Villigen PSI, Switzerland}

\author{F.~Yokaichiya}
\affiliation{Institute for Quantum Phenomena in Novel Materials, Helmholtz-Zentrum Berlin\\ f\"ur Materialen und Energie GmbH, Hahn-Meitner-Platz~1, 14109 Berlin, Germany}

\author{A.~S.~Sukhanov}
\affiliation{Institut f\"ur Festk\"orper- und Materialphysik, Technische Universit\"at Dresden, 01069 Dresden, Germany}
\affiliation{Max Planck Institute for Chemical Physics of Solids, N\"othnitzer Str. 40, 01187 Dresden, Germany}

\author{R.~M.~Eremina}
\affiliation{Zavoisky Physical-Technical Institute, FRC Kazan Scientific Center of RAS, Sibirsky tract 10/7, 420029 Kazan, Russia}

\author{Ph.~Schlender}
\affiliation{Fakult\"at Chemie und Lebensmittelchemie, Technische Universit\"at Dresden, 01069 Dresden, Germany}

\author{A.~Schneidewind}
\affiliation{J\"ulich Center for Neutron Science at MLZ, Forschungszentrum J\"ulich GmbH, Lichtenbergstra{\ss}e 1, 85748 Garching, Germany}

\author{H.-H.~Klauss}
\affiliation{Institut f\"ur Festk\"orper- und Materialphysik, Technische Universit\"at Dresden, 01069 Dresden, Germany}

\author{D.~S.~Inosov}\email[Corresponding author: ]{Dmytro.Inosov@tu-dresden.de}
\affiliation{Institut f\"ur Festk\"orper- und Materialphysik, Technische Universit\"at Dresden, 01069 Dresden, Germany}
\affiliation{W\"urzburg-Dresden Cluster of Excellence on Complexity and Topology in Quantum Matter\,---\,\textit{ct.qmat}, TU Dresden, 01069 Dresden, Germany}

\begin{abstract}\parfillskip=0pt\relax
\noindent The quantum spin systems Cu$_2M^\prime$BO$_5$ (\mbox{$M^\prime$\,=~Al, Ga}) with the ludwigite crystal structure consist of a structurally ordered Cu$^{2+}$ sublattice in the form of three-leg ladders, interpenetrated by a structurally disordered sublattice with a statistically random site occupation by magnetic Cu$^{2+}$ and nonmagnetic Ga$^{3+}$ or Al$^{3+}$ ions. A microscopic analysis based on density-functional-theory calculations for Cu$_2$GaBO$_5$ reveals a frustrated quasi-two-dimensional spin model featuring five inequivalent antiferromagnetic exchanges. A broad low-temperature $^{11\kern-.5pt}$B nuclear magnetic resonance points to a considerable spin disorder in the system. In zero magnetic field, antiferromagnetic order sets in below $T_\text{N}\approx4.1$~K and $\sim$2.4~K for the Ga and Al compounds, respectively. From neutron diffraction, we find that the magnetic propagation vector in Cu$_2$GaBO$_5$ is commensurate and lies on the Brillouin-zone boundary in the $(H0L)$ plane, $q_\text{m} = (0.45,\,0,\,-0.7)$, corresponding to a complex noncollinear long-range ordered structure with a large magnetic unit cell. Muon spin relaxation is monotonic, consisting of a fast static component typical for complex noncollinear spin systems and a slow dynamic component originating from the relaxation on low-energy spin fluctuations. Gapless spin dynamics in the form of a diffuse quasielastic peak is also evidenced by inelastic neutron scattering. Most remarkably, application of a magnetic field above 1~T destroys the static long-range order, which is manifested in the gradual broadening of the magnetic Bragg peaks. We argue that such a crossover from a magnetically long-range ordered state to a spin-glass regime may result from orphan spins on the structurally disordered magnetic sublattice, which are polarized in magnetic field and thus act as a tuning knob for field-controlled magnetic disorder.
\end{abstract}

\pacs{75.25.-j, 75.30.Ds, 28.20.Cz\vspace{-3pt}}

\maketitle

\section{Introduction}

\subsection{General motivation}

\noindent Low-dimensional copper compounds~\cite{SchollwockRichter04, VasilievVolkova19}, among them a number of naturally occurring minerals~\cite{Inosov18}, often feature complex magnetic phase diagrams with multiple competing phases. This competition among different ground states results from a combination of strong quantum fluctuations (due to the low spin $S=1/2$ of the Cu$^{2+}$ ion), low dimensionality, magnetic frustration, and in some cases the effects of disorder\,---\,factors that are all known to weaken or suppress simple collinear magnetic order. As a consequence, in contrast to conventional classical magnets where the ground state is stable against weak external influences such as pressure, strain, or magnetic field, frustrated low-dimensional quantum magnets are highly susceptible to these external factors, so that multiple phase transitions can occur in the experimentally accessible range of control parameters. One of the best studied examples with such behavior is the frustrated spin-chain compound linarite, PbCuSO$_4$(OH)$_2$, in which four distinct magnetic phases were found upon varying the temperature, magnetic field, and its direction~\mbox{\cite{WillenbergSchaepers12, FengPovarov18, HeinzeBastien19}}. The zero-field spin-spiral ground state is suppressed already in a moderate magnetic field of $\sim$2.5~T, resulting in a field-driven quantum phase transition. Different types of field-induced quantum critical points were found in many other low-dimensional spin-1/2 compounds with magnetic frustration~\cite{ParkChoi07, RuleTennant11, KonoSakakibara15, PovarovBhartiya19}.

Ludwigite oxyborates with the general chemical formula $M_2^{2+}M^{\prime{\kern.5pt}3+}$BO$_5$ represent a family of compounds that derive from the crystal structure of the natural magnesium-iron borate mineral ludwigite, Mg$_2$FeBO$_5$. From the point of view of quantum magnetism, synthetic compounds where the divalent metal ion $M$ is represented by magnetic Cu$^{2+}$ with $S=1/2$ attract most attention. For simplicity, the trivalent metal ion $M^\prime$ should preferably remain nonmagnetic (for instance, Al$^{3+}$ or Ga$^{3+}$)~\cite{Schaefer1995, Hriljac1990, PetrakovskiiBezmaternykh09}, although ludwigites with magnetic trivalent ions, e.g. $M^\prime$~=~Fe$^{3+}$ ($S=5/2$), Mn$^{3+}$ ($S=2$), or Cr$^{3+}$ ($S=3/2$), are also well known~\mbox{\cite{PetrakovskiiBezmaternykh09, SofronovaNazarenko16, MoshkinaRitter17, MoshkinaPlatunov18, SottmannNataf18}}. In particular, some of them find practical applications as conversion-type electrode materials for lithium-ion batteries~\cite{SottmannNataf18, PralongRoux17}. It has been suggested that the ludwigite structure is closely related to that of the common aristotype of malachite and rosasite~\cite{GirgsdiesBehrens12}. However, the lattice symmetry of the copper ludwigites Cu$_2$GaBO$_5$ and Cu$_2$AlBO$_5$ is in reality lowered to monoclinic because of Jahn-Teller distortions of the CuO$_6$ octahedra~\cite{PetrakovskiiBezmaternykh09}.

Initial low-temperature characterization measurements revealed relatively sharp phase transitions in the magnetic susceptibility and specific heat to a presumably antiferromagnetic (AFM) state with the N\'eel temperature, $T_\text{N}$, between 3.4 and 4.1~K in Cu$_2$GaBO$_5$~\cite{PetrakovskiiBezmaternykh09, EreminaGavrilova20} and $\sim$2.4~K in Cu$_2$AlBO$_5$~\cite{EreminaGavrilova20}. Surprisingly, the application of magnetic field does not drive this transition towards a quantum critical point, as in the majority of other low-dimensional quantum magnets, but suppressed it by smearing it out. For instance, a sharp cusp observed in specific heat in zero field transforms into a broadened hump in fields above 1~T~\cite{EreminaGavrilova20}. As we are going to demonstrate in the following, this crossover is accompanied by a drastic reduction in the magnetic correlation length and can be therefore seen as a field-induced transition from long-range magnetic order to a highly disordered spin-glass state with only short-range spin correlations. While it is generally expected that the presence of structural disorder on the magnetic sublattice should lead to such glassy states, and they have been reported previously in other ludwigite-type compounds~\cite{FermandesGuimaraes98, ContinentinoFernandes99, KnyazevIvanova12, MoshkinaSofronova16}, it is remarkable that the disorder is inactive in zero field and can be ``activated'' by the application of a moderate magnetic field of the order of only several teslas, giving us a tuning knob to change the degree of magnetic disorder and the magnetic correlation length in the system continuously. These considerations motivated our present study to reveal the details and explain the mechanism of the field-driven destruction of long-range magnetic order in favor of a spin-glass-like state in copper ludwigites.

\vspace{-3pt}\subsection{\label{sec:str}Crystal structure and known magnetic properties}\vspace{-3pt}

\begin{figure}[t!]
\includegraphics[width=\columnwidth]{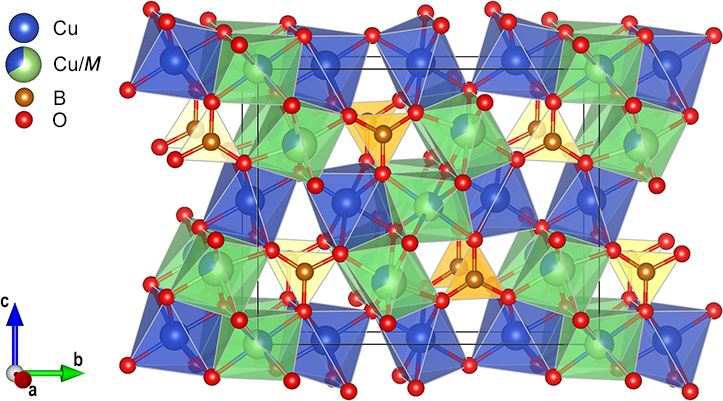}\vspace{-1pt}
\caption{Crystal structure of Cu$_2M$BO$_5$ ludwigites, where $M$ is either Ga~\cite{Schaefer1995} or Al~\cite{Hriljac1990}. The magnetic Cu$^{2+}$ ions occupy four inequivalent positions in distorted octahedral coordinations, which form edge-sharing zigzag walls perpendicular to the $\mathbf{b}$ axis. The chains of edge-sharing octahedra surrounding the $M(1)$ and $M(2)$ sites with 100\% occupation by the Cu$^{2+}$ ions in neighboring zigzag walls (blue) are connected triplewise by corner sharing into three-leg ladders running along the $\mathbf{a}$ axis. They are separated by the structurally disordered three-leg ladders ($3\times\infty$ ribbons) of edge-sharing octahedra (green) surrounding the $M(3)$ and $M(4)$ sites that are statistically occupied by magnetic Cu$^{2+}$ and nonmagnetic $M^{3+}$ ions. Solid lines mark the unit cell. The visualization was done in VESTA~\cite{MommaIzumi11}.\vspace{-2pt}}
\label{Fig:CrystStruct}
\end{figure}

In the present paper, we report the results of magnetic measurements on the two copper ludwigites Cu$_2$GaBO$_5$ and Cu$_2$AlBO$_5$ that share the crystal structure depicted in Fig.~\ref{Fig:CrystStruct}, according to the available x-ray structure refinement~\cite{Schaefer1995, Hriljac1990} and our own refinement based on the results of single-crystal x-ray diffraction. The unit cell is monoclinic, described by the space group $P2_1/c$ with the lattice parameters $a=3.1126(3)$~\AA, $b=11.9215(13)$~\AA, $c=9.4792(10)$~\AA, and $\beta=97.909(9)^\circ$ (for Cu$_2$GaBO$_5$) or $a=3.0633(4)$~\AA, $b=11.7744(17)$~\AA, $c=9.3537(13)$~\AA, and $\beta=97.721(11)^\circ$ (for Cu$_2$AlBO$_5$) at room temperature. There are in total four structurally inequivalent metal sites: $M(1)$ in Wyckoff position $4e$, $M(2)$ in Wyckoff position $2d$, $M(3)$ in Wyckoff position $2a$, and $M(4)$ in Wyckoff position $4e$. Two of them, namely $M(1)$ and $M(2)$, are fully occupied with Cu$^{2+}$, whereas the other two, $M(3)$ and $M(4)$, have fractional occupation with Cu$^{2+}$ and the nonmagnetic $M^{\prime{\kern.5pt}3+}$ ion in the ratio close to Cu:$M^\prime$=1:2. This results in the stoichiometric chemical composition of the compound within the accuracy of elemental composition analysis. The fact that a stoichiometric compound is formed in spite of the statistical occupation of metal sites could be an indication that the true equilibrium crystal structure has a tendency to the formation of a superstructure on the disordered metal sublattice, yet no evidence for such a superstructure has been found in single-crystal x-ray or neutron diffraction data up to now, to the best of our knowledge.

\begin{figure}[b!]
	\includegraphics[width=\linewidth]{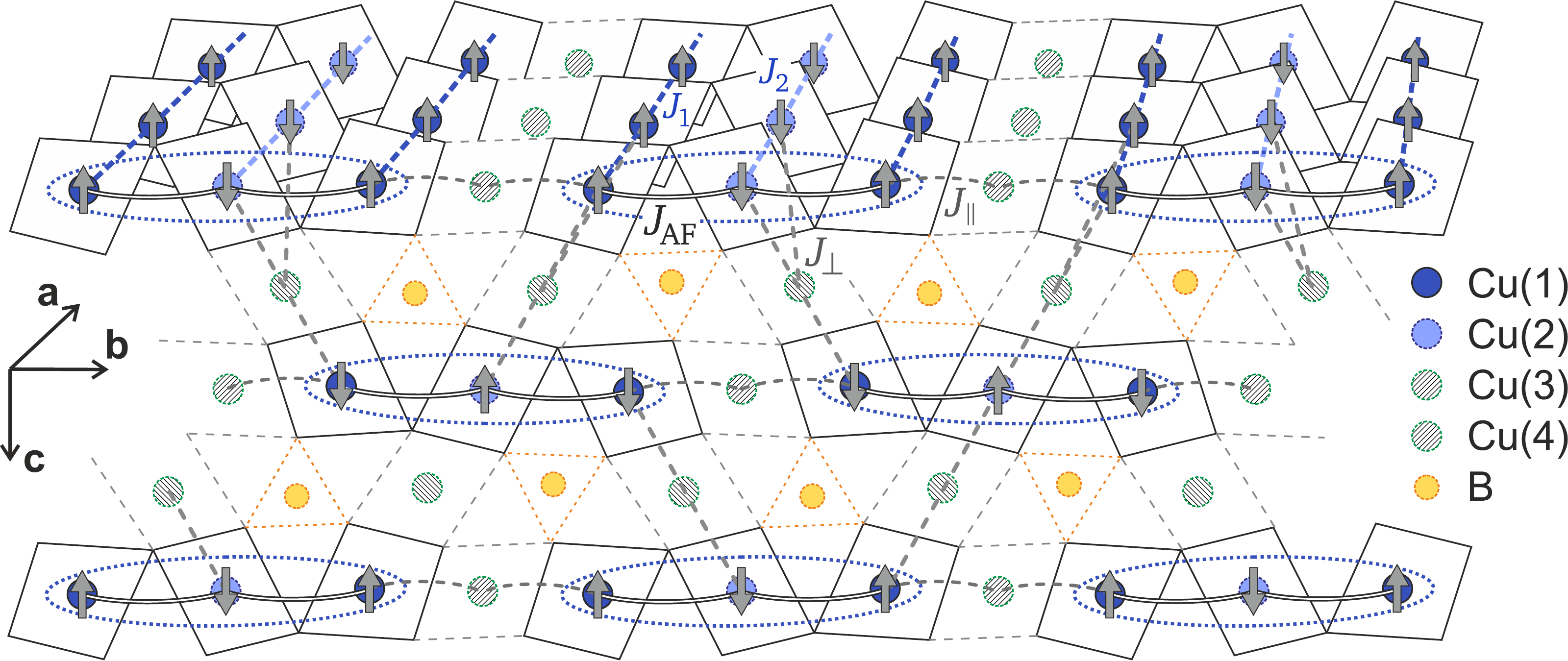}
	\caption{Sketch of the crystal structure with the hypothetical magnetic order that one might expected in the simplified three-leg ladder model suggested by the GKA rules, but which is not observed in reality. Nearest-neighbor interactions between the Cu$^{2+}$ ions are shown by solid double lines (AFM) and dashed lines (FM), respectively. Hatched circles show the fractionally occupied metal sites. Dotted ellipses highlight the AFM trimers. To simplify the drawing, spin chains along the $\mathbf{a}$ axis are shown for the top row of trimers only.\vspace{-4pt}}
\label{Fig:Ladders}
\end{figure}

\begin{figure*}[t]
\includegraphics[width=\linewidth]{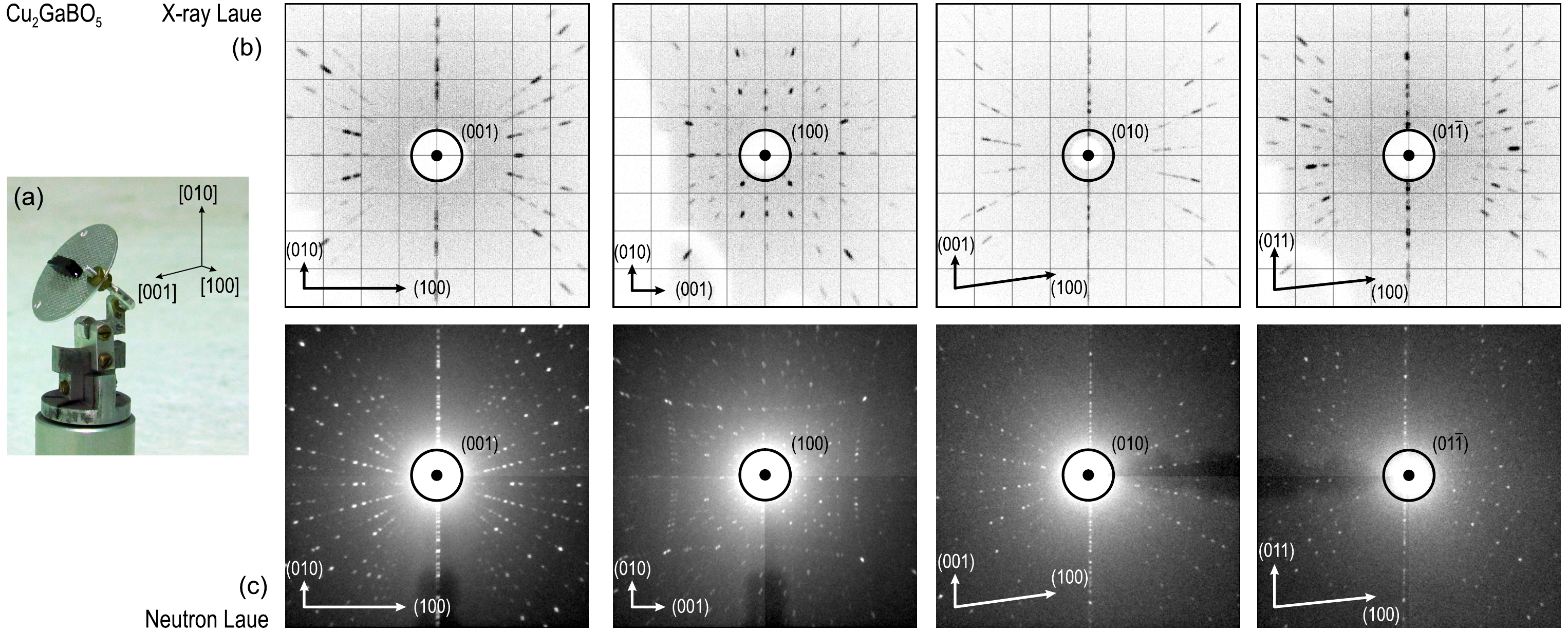}
\caption{(a)~Oriented Cu$_2$GaBO$_5$ single crystal on an aluminum sample holder. (b)~X-ray and (c)~neutron Laue diffraction measurements for different crystal directions at room temperature.}
\label{Fig:Laue}
\end{figure*}

All four metal sites in the ludwigite lattice are coordinated with distorted oxygen octahedra that are highlighted in Fig.~\ref{Fig:CrystStruct} with blue and green shading for the fully and fractionally occupied Cu sites, respectively. The former are connected by corner sharing along the $\mathbf{b}$ direction, forming Cu(1)--Cu(2)--Cu(1) trimers with the Cu--O--Cu bridging angle of 119.1$^\circ$ ($M^\prime$~=~Ga) or 117.3$^\circ$ ($M^\prime$~=~Al). According to the empirical Goodenough-Kanamori-Anderson (GKA) rules~\cite{GKA-Rules}, this should correspond to a strong AFM superexchange interaction within the trimer. Along the $\mathbf{a}$ direction, trimers are connected into three-leg ladders by edge sharing, which results in the bridging angles slightly smaller than 90$^\circ$. For the $M^\prime$~=~Ga compound, they constitute 89.4$^\circ$ along the Cu(2)--O--Cu(2) bonds that form the central leg of the ladder and 87.3$^\circ$ for the Cu(1)--O--Cu(1) outer bonds (in the Al compound, the angles are similar). This should result in a ferromagnetic (FM) superexchange along the ladder. Such three-leg spin ladders with an AFM coupling along the ``rungs'' and two different FM couplings along the ``legs'' form the main building blocks that are at first glance expected to define the magnetic behavior of copper ludwigites. It is worth noting that spin ladders with an opposite sign of superexchange interactions along the rungs and legs are quite unusual among low-dimensional cuprates. For comparison, the most studied families of spin-ladder compounds Sr$_{n-1}$Cu$_{n+1}$O$_{2n}$~\cite{KojimaKeren95, ThurberImai00} and La$_{4+4n}$Cu$_{8+8n}$O$_{14+8n}$~\cite{CavaSiegrist91, ProkesRessouche17} have nearly identical 180$^\circ$ Cu--O--Cu bonds in both directions, which results in AFM interactions of similar magnitude both in the rung and leg directions.

In this na\"{\i}ve model, the magnetic interladder coupling is mediated by the Cu$^{2+}$ ions that statistically populate 1/3 of the $M^\prime$(3) and $M^\prime$(4) sites. They are surrounded by edge-sharing oxygen octahedra, forming $3\times\infty$ ribbons with nearly perpendicular Cu--O--Cu bonds. It is therefore expected that this dilute, structurally disordered sublattice of Cu$^{2+}$ orphan spins should provide, on average, a weak FM coupling between the neighboring trimers. In particular, the Cu(3) sites connect two outer spins in adjacent trimers along the $\mathbf{b}$ axis, whereas Cu(4) sites couple the outer spin of one trimer to the central spin of a neighboring trimer along the $\mathbf{c}$ axis. This should result in a dilute random network of weak FM bonds between the three-leg ladders. This oversimplified picture, resulting from the application of GKA rules to the nearest-neighbor Cu--O--Cu bonds only, is schematically illustrated in Fig.~\ref{Fig:Ladders}. The mentioned interactions are not frustrated and are expected to result in a collinear $\mathbf{q}=0$ type of AFM order depicted in Fig.~\ref{Fig:Ladders}, contrary to the experimental observations.

From the experimental point of view, the Curie-Weiss temperatures of approximately $-70$ and $-50$~K for the Ga and Al compounds, respectively~\cite{EreminaGavrilova20}, are about 20 times higher by absolute value than the N\'eel temperature, which suggests relatively strong frustration of superexchange interactions. The neutron-diffraction data presented below in section \ref{Sec:NeutronDiffraction} also indicate that the AFM order in Cu$_2$GaBO$_5$ is noncollinear and resembles a spin spiral propagating in the $(H0L)$ plane, which is clearly inconsistent with the simplified model presented above. This inconsistency motivated us to perform more rigorous first-principles calculations of the superexchange interactions, presented in section \ref{Sec:Theory}, which resulted in a more accurate magnetic model of copper ludwigites.

\vspace{-3pt}\section{Sample preparation and characterization}\vspace{-2pt}

Large translucent dark-green Cu$_2M^{\prime\,11\kern-.8pt}$BO$_5$ single crystals ($M^{\prime}$~=~Ga,\,Al) have been grown by the flux method as described elsewhere~\cite{EreminaGavrilova20}. For the purposes of our study, which includes neutron scattering, we used isotopically enriched boric acid with 99.88~at.\,\% $^{11\kern-.8pt}$B isotope from Ceradyne, Inc. as a starting material to minimize neutron absorption by the $^{10\kern-.5pt}$B isotope. The same isotope-enriched samples were also beneficial for $^{11\kern-.8pt}$B nuclear magnetic resonance (NMR) measurements, whereas all other measurements presented here were performed on the same set of samples for consistency. We have investigated both Cu$_2$GaBO$_5$ and Cu$_2$AlBO$_5$ compounds using muon spin relaxation ($\mu$SR). In view of their similar properties, however, some other measurements (neutron scattering, NMR) were restricted only to the Ga compound for time-saving reasons.

The single-crystal samples were first of all characterized by x-ray diffraction at ambient temperature, which revealed no deviations from the previously published crystal structure~\cite{Schaefer1995, Hriljac1990}, as well as by magnetic susceptibility, specific heat, and magnetization measurements in fields up to 9~T. These results are summarized in Ref.~\cite{EreminaGavrilova20}. To ensure good crystallinity of our samples and to orient them before the neutron-scattering measurements, we also collected x-ray and neutron Laue diffraction patterns on some of our single crystals. For the x-ray Laue measurements, we used the in-house MWL~120 real-time back-reflection Laue camera system from Multiwire Laboratories, Ltd., whereas neutron Laue diffraction measurements were performed at the E11 Fast Acquisition Laue Camera for Neutrons (FALCON) at HZB, Berlin. The corresponding results for one of the largest Cu$_2$Ga$^{11\kern-.8pt}$BO$_5$ single crystals, viewed from various high-symmetry crystal directions, are shown in Fig.~\ref{Fig:Laue}\,(b,\,c). The oriented crystal, mounted on an aluminum sample holder in the $(H0L)$ scattering plane for the low-temperature neutron-diffraction measurements, is shown in Fig.~\ref{Fig:Laue}\,(a).

\vspace{-3pt}\section{Experimental methods and results}\vspace{-2pt}

\subsection{Single-crystal neutron diffraction}\vspace{-3pt}
\label{Sec:NeutronDiffraction}

\begin{figure*}
\includegraphics[width=\linewidth]{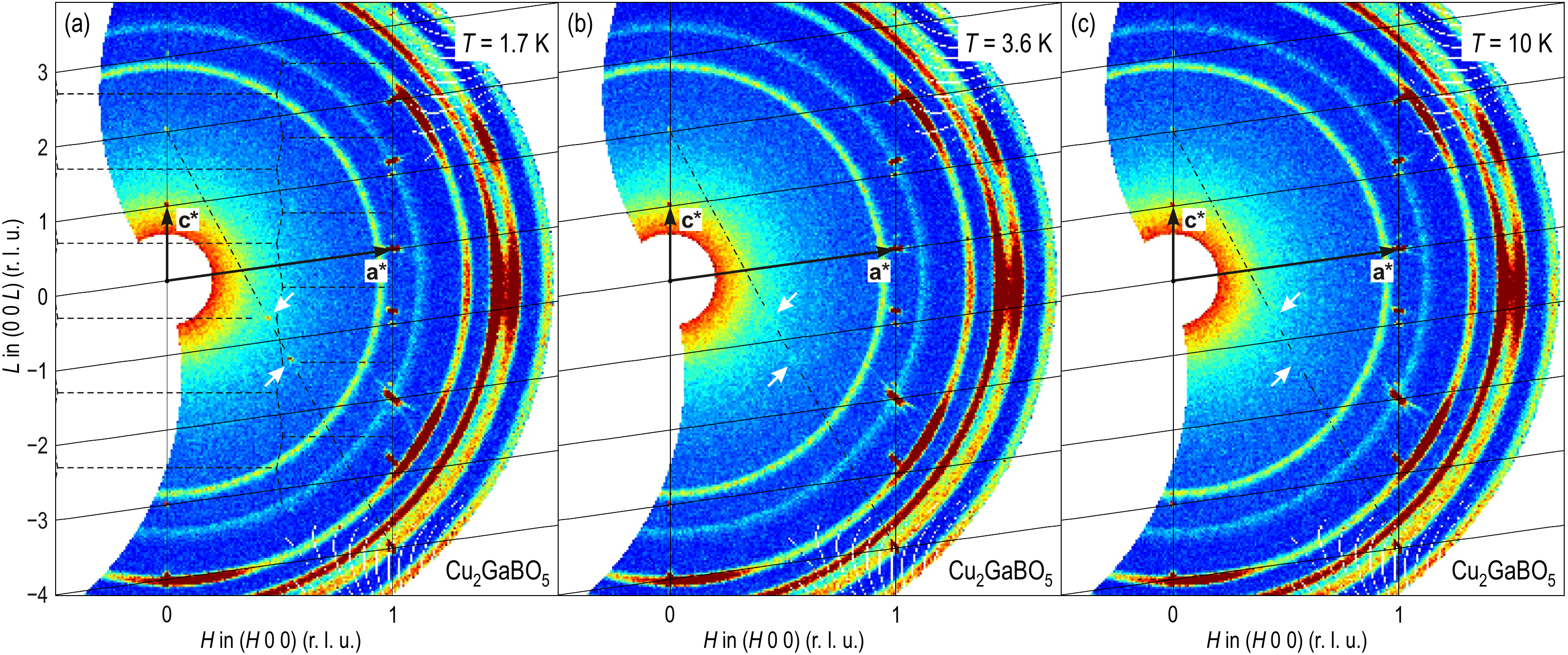}
\caption{Neutron diffraction data, measured at the E2 single-crystal diffractometer at HZB on one of the largest single crystals of the Cu$_2$Ga$^{11\kern-.8pt}$BO$_5$ ludwigite. Magnetic Bragg peaks at the AFM propagation vector $\mathbf{q}_\text{m}=(0.45~0~-\!0.7)$ and the equivalent position \mbox{$({\kern-0.5pt}10\overline{2})-\mathbf{q}_\text{m}$} are shown with white arrows. Note that both peaks fall on the intersection of the Brillouin-zone boundary (dashed lines) with the straight dash-dotted line connecting the $(002)$ and $({\kern-0.5pt}10\overline{4})$ reciprocal-lattice vectors.}
\label{Fig:DiffractionMaps}
\end{figure*}

To reveal the magnetic propagation vector in the AFM state, we measured the single crystal of Cu$_2$GaBO$_5$ depicted in Fig.~\ref{Fig:Laue}\,(a) by single-crystal neutron diffraction using the Flat-Cone Diffractometer E2 at HZB, Berlin. Neutrons with the wavelength $\lambda=2.41$~\AA\ were selected from the thermal neutron beam with a PG$(002)$ monochromator. A pyrolytic graphite (PG) filter was used to reduce the higher-order contamination from the monochromator. The crystal was mounted in the $(H0L)$ scattering plane, and the data were collected at various temperatures between 1.7 and 10~K and magnetic fields from 0 to 6~T applied along the $\mathbf{b}$ axis. Complete reciprocal-space maps at 1.7, 3.6, and 10~K in zero magnetic field are shown in Fig.~\ref{Fig:DiffractionMaps}.

The ringlike features in the color maps originate from powder scattering on aluminum of the sample holder and the cryomagnet. Bragg reflections that fall on the intersections of coordinate axes originate from the oriented sample. Much weaker Bragg peaks originating from the second mono\-clinic twin domain present in the same single crystal appear reflected about the horizontal axis, resulting in sharp intensity spots offset vertically from the reciprocal-lattice vectors. In the largest single crystal used in our neutron-diffraction measurements, the intensity of Bragg peaks originating from the minority twin was about three times lower compared to the identical reflections from the main twin. Moreover, second-order Bragg scattering from the monochromator, even after being weakened by the PG filter, is still present to some extent in the incident neutron beam. This leads to the appearance of weak $(H0L)$ Bragg reflections at structurally forbidden positions with odd $L$, e.g. $(001)$ or $(10\overline{1})$, whenever the corresponding $(2H~0~2L)$ reflection has high intensity.

\begin{figure}[b!]\vspace{-4pt}
\centerline{\includegraphics[width=0.8\columnwidth]{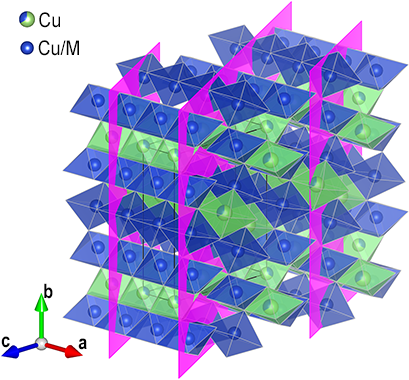}}\vspace{-4pt}
\caption{A set of equidistant planes in direct space, showing the pitch of the AFM spin-spiral structure in Cu$_2$Ga$^{11\kern-.8pt}$BO$_5$. The planes of equal phase are orthogonal to the magnetic propagation vector $\mathbf{q}_\text{m}$ and are spaced by $2\pi/|\mathbf{q}_\text{m}|$. In the crystal structure, only the distorted octahedra around the magnetic Cu ions are shown for simplicity. The visualization was done in VESTA~\cite{MommaIzumi11}.\vspace{-4pt}}
\label{Fig:CrystalPlanes}
\end{figure}

Magnetic Bragg peaks can be seen in the $T=1.7$~K dataset in Fig.~\ref{Fig:DiffractionMaps} at the \mbox{$(0.45~0~-\!0.7)$} and \mbox{$(0.55~0~-\!1.3)$} wave vectors, as indicated by the white arrows. The shortest of these wave vectors corresponds to the magnetic propagation vector \mbox{$\mathbf{q}_\text{m}=(0.45~0~-\!0.7)$}, while the other one represents the magnetic satellite of the strong $({\kern-0.5pt}10\overline{2})$ structural Bragg reflection, located at \mbox{$({\kern-0.5pt}10\overline{2})-\mathbf{q}_\text{m}$}. Weaker magnetic satellites can be also recognized at $(0.55~0~0.7)=({\kern-0.5pt}100)-\mathbf{q}_\text{m}$ and $(0.55~0~-3.3)=({\kern-0.5pt}10\overline{4})-\mathbf{q}_\text{m}$. Magnetic reflections from the minority twin domain cannot be seen because of their low intensity, with the exception of the strongest magnetic Bragg peak at \mbox{$({\kern-0.5pt}10\overline{2})-\mathbf{q}_\text{m}$}, which nearly coincides with the weak $({\kern-0.5pt}100)-\mathbf{q}_\text{m}$ reflection from the main twin. As a result, the peak at $(0.55~0~0.7)$ appears slightly split. All the mentioned reflections weaken with increasing temperature and disappear above $T_{\rm N}$, see Figs.~\ref{Fig:DiffractionMaps}\,(b,\,c), which confirms their magnetic origin.

\begin{figure*}[t!]\vspace{-2pt}
\includegraphics[width=\linewidth,height=1.3\columnwidth]{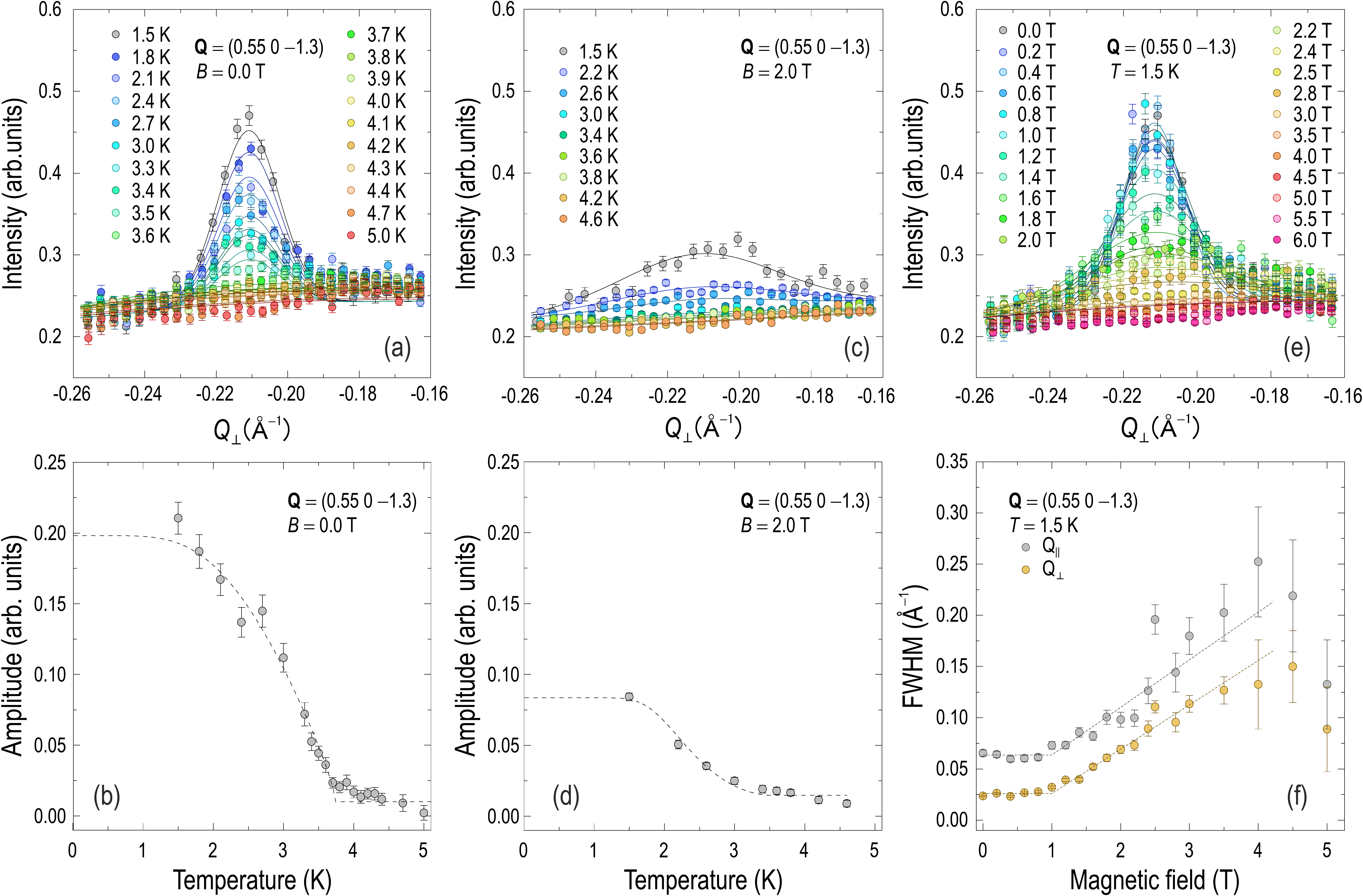}
\caption{Magnetic-field and temperature dependence of the $(0.55~0~-\!1.3)$ magnetic Bragg peak from the E2 data. (a,\,c)~Neutron diffraction data, measured at various temperatures in zero magnetic field and at 2~T, respectively. (b,\,d)~The corresponding $T$-dependence of the magnetic Bragg intensity (after background subtraction) at zero field and at 2~T. (e)~Magnetic-field dependence of the intensity profile across the magnetic Bragg peak at the base temperature, $T=1.5$~K. (f)~Magnetic-field dependence of the full width at half maximum of the peaks shown in panel (e). All the data points in panels (b), (d), and (f) were obtained from Gaussian fits that are shown in panel (a), (c), and (d), respectively, with solid lines.}
\label{Fig:E2_peaks}	
\end{figure*}

The two strongest magnetic Bragg reflections can be seen as satellites around the commensurate $(\frac{1}{2}0\overline{1})$ wave vector at the zone boundary, suggesting that the magnetic structure represents some sort of a spin spiral with nearly AFM spin arrangement along the $\mathbf{a}$ axis (spin-ladder direction). This structure is in addition twisted into a spiral, as illustrated in Fig.~\ref{Fig:CrystalPlanes} by a set of equidistant planes of equal phase that are orthogonal to the propagation vector and are placed at a distance $2\pi/|\mathbf{q}_\text{m}|$ from each other. The absence of higher-order reflections in the diffraction data suggests that this twisting is to a good approximation uniform within the unit cell. Clearly, such a structure cannot result from the na\"{i}ve magnetic model shown in Fig.~\ref{Fig:Ladders}, which implies the spin arrangement along the $\mathbf{a}$ axis to be ferromagnetic.

A very distinctive feature of the experimentally observed magnetic structure, however, is that it appears to be commensurate within the accuracy of our measurements. The two strongest magnetic satellites at \mbox{$(0.45~0~-\!0.7)$} and \mbox{$(0.55~0~-\!1.3)$} fall onto a straight diagonal line connecting the $(002)$ and $(10\overline{4})$ structural Bragg reflections (dash-dotted line in Fig.~\ref{Fig:DiffractionMaps}), and the distance between them constitutes exactly 1/10 of the length of the $(10\overline{6})$ vector, suggesting that this magnetic structure represents a complex commensurate AFM order with a large magnetic unit cell and the propagation vector of precisely \mbox{$\left(\frac{9}{20}~0~\frac{-7}{10}\right)$}. Quite surprisingly, within about 2\% accuracy this propagation vector falls on the Brillouin-zone boundary (dashed lines in Fig.~\ref{Fig:DiffractionMaps}), which suggests a possible geometric constraint imposed on the lattice parameters by the inner structure of the crystallographic unit cell. The mechanism behind such a lock-in of the AFM spin spiral to this commensurate vector with a large denominator remains unclear, and it is evident that the full magnetic refinement of such a complex magnetic structure is not feasible, based on just a few experimentally observable magnetic Bragg reflections.

We now proceed with a more detailed analysis of the strongest magnetic Bragg peak at \mbox{$(0.55~0~-\!1.3)$} in an external magnetic field, applied along the $\mathbf{b}$ axis. The zero-field data are shown in Figs.~\ref{Fig:E2_peaks}\,(a,\,b). The intensity profiles at various temperatures were obtained by integrating the E2 data from the vicinity of the magnetic Bragg reflection in the longitudinal direction and plotting them vs. the transverse momentum component $Q_\perp$. The resulting curves were fitted with Gaussian peak functions, as shown in Fig.~\ref{Fig:E2_peaks}~(a) with solid lines. The temperature dependence of the peak amplitude, resulting from these fits, is presented in Fig.~\ref{Fig:E2_peaks}~(b), where it has been fitted with an order-parameter model (dashed line). One can see that the magnetic transition in zero field is quite sharp, with a $T_{\rm N}$ of $\sim$3.8~K. However, the application of a magnetic field of only 2~T changes the situation qualitatively. The corresponding datasets in Figs.~\ref{Fig:E2_peaks}\,(c,\,d) show a twice broader peak, and the corresponding temperature dependence evidences a smeared phase transition with a smooth intensity onset. A similar field-induced broadening of the phase transition was also observed in the specific-heat data from the same samples~\cite{EreminaGavrilova20}.

\begin{figure*}[t!]\vspace{-3pt}
\centerline{\includegraphics[width=0.7\linewidth]{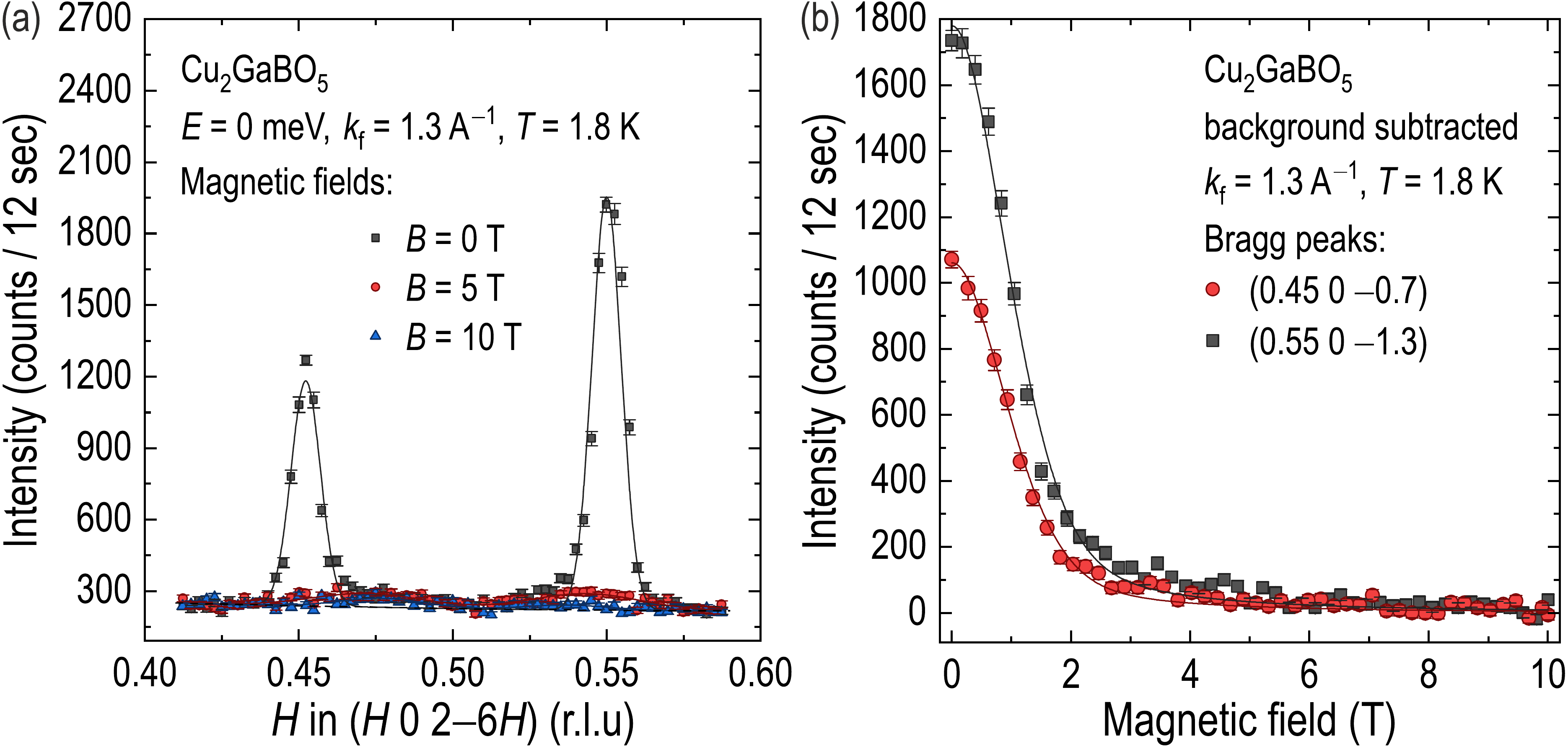}}\vspace{-2pt}
\caption{Suppression of the magnetic Bragg peaks in Cu$_2$GaBO$_5$ by an external magnetic field. (a)~Elastic neutron-scattering intensity along the $(H~0~2\!-\!6H)$ line in momentum space, measured in zero field and in the field of 5 and 10~T, applied along the $\mathbf{b}$ axis. (b)~Magnetic-field dependence of the peak intensity, measured on top of the two Bragg reflections at $(0.45~0~-\!0.7)$ and $(0.55~0~-\!1.3)$.}
\label{Fig:NeutronElastic}
\end{figure*}

\begin{figure*}[t!]\vspace{-1pt}
\centerline{\hspace{0.95em}\includegraphics[width=0.698\linewidth]{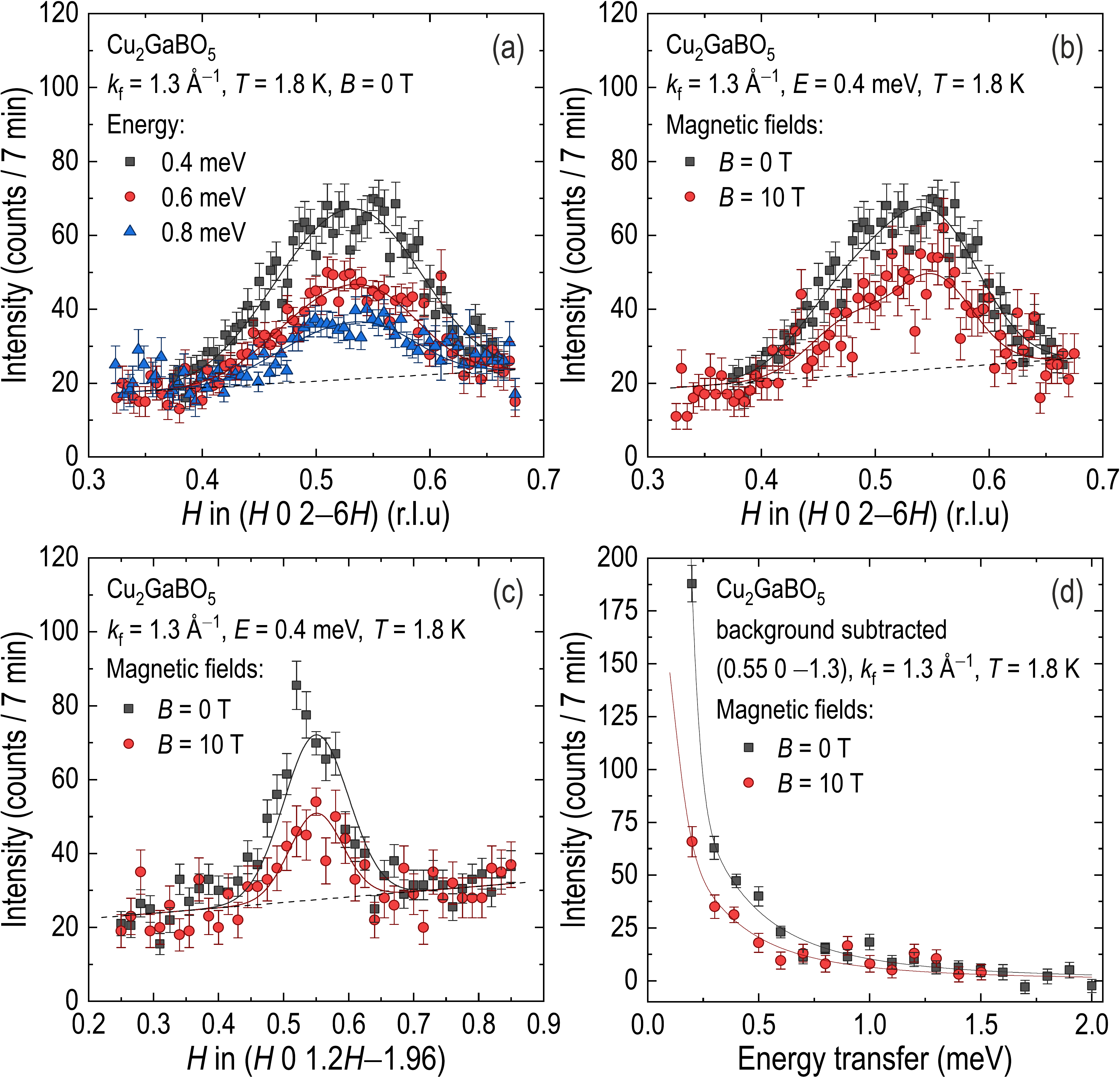}}\vspace{-2pt}
\caption{Inelastic neutron scattering data collected at the \textsc{Panda} spectrometer at the base temperature of 1.8~K. (a)~Constant-energy scans along the $(H~0~2\!-\!6H)$ line in momentum space, measured in zero magnetic field at several energies. The solid lines are fits with two Gaussian peaks of different amplitude, symmetrically offset from the commensurate $(\frac{1}{2}\,0\,\overline{1})$ wave vector, plus a linear background (shown by a dashed line). (b)~Comparison of the $\mathbf{Q}$-scans at an energy transfer of 0.4~meV, measured in zero field and in a 10~T field applied along the $\mathbf{b}$ axis. (c)~Corresponding scans in the orthogonal $\mathbf{Q}$ direction, $(H~0~1.2H\!-\!1.96)$. (d)~Energy dependence of the background-subtracted magnetic intensity in zero and 10~T magnetic field. Solid lines are guides to the eyes.\vspace{-1pt}}
\label{Fig:NeutronInelastic}
\end{figure*}

The dependence of the full width at half maximum (FWHM) of the same reflection on magnetic field, measured at the base temperature of 1.5~K, is presented in Figs.~\ref{Fig:E2_peaks}\,(e,\,f). The Bragg peaks are elongated in the longitudinal direction (along $\mathbf{Q}$), as one can see from the comparison of longitudinal and transverse peak widths in Figs.~\ref{Fig:E2_peaks}\,(f). In zero magnetic field, the magnetic Bragg reflection is essentially resolution limited. The lower limit for the magnetic correlation length, estimated from the peak width, is approximately 250~\AA. With the application of magnetic field perpendicular to the scattering plane, the peak remains sharp within the experimental resolution below 1~T, but then starts to broaden continuously, which implies a gradual reduction in the magnetic correlation length and a destruction of the long-range magnetic order in favor of a spin-glass-like state, in agreement with the specific-heat data~\cite{EreminaGavrilova20}.\enlargethispage{2pt} The fitting results above 3~T are no longer accurate, as the peak becomes very weak and extends outside of our limited scanning range in the rocking-angle direction.

The suppression of magnetic Bragg reflections in an external magnetic field has also been confirmed in the elastic neutron scattering measurements presented in Fig.~\ref{Fig:NeutronElastic}. These data were collected at the cold-neutron triple axis spectrometer \textsc{Panda} (Heinz Maier-Leibnitz Zentrum, Garching, Germany)~\cite{SchneidewindLink06, SchneidewindCermak15} with the final neutron wave vector fixed at $k_\text{f}=1.3$~\AA$^{-1}$. This time, the sample represented a mosaic of many coaligned single crystals of Cu$_2$GaBO$_5$ with a total mass of 1.8~g and a mosaicity of $\sim\!1^\circ$, glued to an Al plate with a small amount of GE varnish. It was mounted in the vertical-field 12~T shielded closed-cycle cryomagnet JVM12 (Oxford Instruments). To suppress higher-order scattering from the monochromator, a cold beryllium filter was installed between the sample and the analyzer. Two magnetic Bragg peaks, centered at $\mathbf{q}_\text{m}=(0.45~0~-\!0.7)$ and $({\kern-0.5pt}10\overline{2})-\mathbf{q}_\text{m}=(0.55~0~-\!1.3)$, are revealed in a scan along the $(H~0~2\!-\!6H)$ direction [black squares in Fig.~\ref{Fig:NeutronElastic}\,(a)]. Note that the second peak has about 60\% higher intensity as a satellite of the very strong structural reflection $({\kern-0.5pt}10\overline{2})$. In a magnetic field applied along the $\mathbf{b}$ axis, the peak amplitude is rapidly suppressed, as shown in Fig.~\ref{Fig:NeutronElastic}\,(b). This cannot be explained by the ferromagnetic polarization of the magnetic structure in an external field, because the saturation field for Cu$_2$GaBO$_5$ exceeds 10~T, judging from the magnetization measurements~\cite{EreminaGavrilova20}. Note that the suppression starts already at small fields below 1~T, even if no broadening of the Bragg reflections can yet be resolved in this field range. At fields above 2~T, only two very broad and weak diffuse-scattering peaks can be recognized in the $\mathbf{Q}$ scan [circles and triangles in Fig.~\ref{Fig:NeutronElastic}\,(a)].

\vspace{-2pt}\subsection{Inelastic neutron scattering}\vspace{-2pt}
\label{Sec:INS}

Using the same spectrometer configuration, we have also measured low-energy spin fluctuations in the vicinity of the two magnetic Bragg peaks by inelastic neutron scattering (INS). Constant-energy scans, presented in Fig.~\ref{Fig:NeutronInelastic}\,(a), show a broad peak around $(0.5~0~-\!1)$. The fact that the center of mass of the peak is shifted to the right, and it is much narrower in the orthogonal direction [Fig.~\ref{Fig:NeutronInelastic}\,(c), black squares], suggests that it actually represents a sum of two unresolved incommensurate peaks centered at the two magnetic satellites, whose intensity ratio follows that of the underlying magnetic Bragg peaks. This assumption was used for the fitting models in Fig.~\ref{Fig:NeutronInelastic}\,(a,\,b), which show good agreement with the data. With an increased energy transfer, the peak intensity is monotonically reduced on an energy scale of about 1~meV, as we recognize from the background-subtracted spectrum in Fig.~\ref{Fig:NeutronInelastic}\,(d).

The application of a magnetic field [Fig.~\ref{Fig:NeutronInelastic}\,(b,\,c), circles] suppresses the intensity of spin fluctuations, yet it falls off much slower than in the elastic channel. The field of 10~T suppresses the inelastic signal by only about a factor of 2, whereas the magnetic Bragg intensity is almost fully suppressed already at 5~T. The two energy spectra of the INS intensity after subtraction of the nonmagnetic background are compared in Fig.~\ref{Fig:NeutronInelastic}\,(d). The magnetic signal falls off monotonically with energy transfer and has no energy structure either in zero field or at the highest measured field of 10~T. Therefore, spin fluctuations can be considered quasielastic, with no evidence for a spin gap down to at least 0.2~meV. This indicates that the magnetocrystalline anisotropy in copper ludwigites must be very low, which is a favorable condition for the formation of noncollinear magnetic order resulting from bond frustration.

According to the total moment sum rule, the suppressed magnetic spectral weight must be transferred to a different point in momentum space. We anticipate that with the application of magnetic field this spectral weight accumulates at the zone center ($\mathbf{q}=0$), judging from the monotonic increase in magnetization in the same field range~\cite{EreminaGavrilova20}. This behavior is analogous to a change in the excitation spectrum of a proper-screw helical spin structure, as it is transformed into a conical screw in a longitudinal magnetic field. A direct verification of this scenario in our cold-neutron INS measurements would not be feasible due to the high nonmagnetic background near the zone center from the tails of the structural Bragg reflections.

\begin{figure*}[t!]
\centerline{\includegraphics[width=0.8\linewidth]{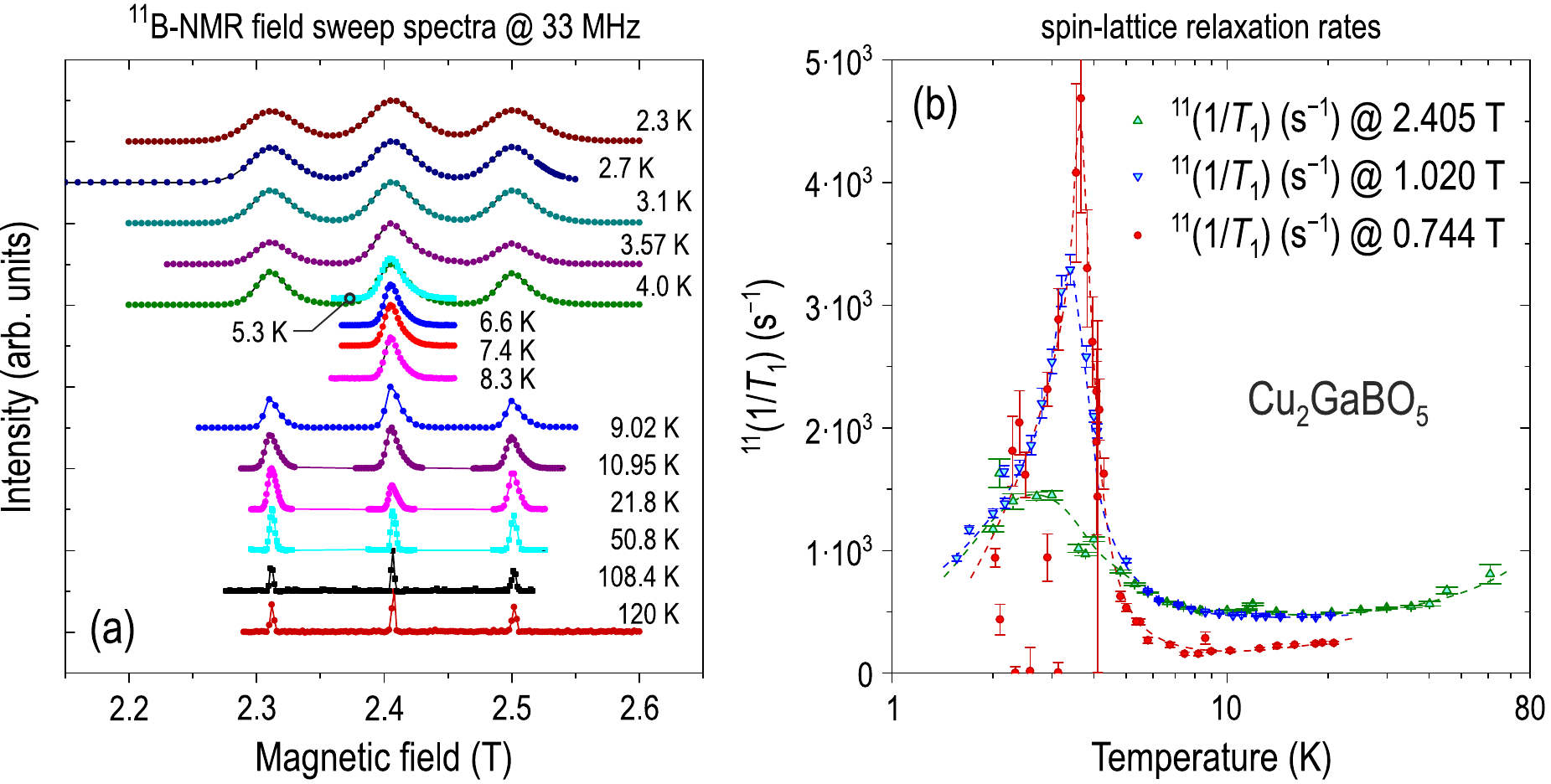}}
\caption{(a)~Field-sweep $^{11\kern-.5pt}$B NMR spectra of in Cu$_2$GaBO$_5$ collected at 33~MHz at various temperatures. (b)~Temperature dependence of the spin-lattice relaxation rates at three different external fields. The dashed lines are guides to the eyes.\vspace{-1pt}}
\label{Fig:NMR}
\end{figure*}

\vspace{-3pt}\subsection{Nuclear magnetic resonance (NMR)}\vspace{-2pt}

Magnetic order and low-energy spin dynamics have been also probed by $^{11\kern-.5pt}$B~NMR, which is a powerful microscopic probe that can shed light on the low-temperature behavior of frustrated magnetic systems, especially in the presence of magnetic disorder. The experiments were performed on the same $^{11\kern-.5pt}$B-enriched single crystals that were used for INS measurements. The $^{11\kern-.5pt}$B isotope has a nuclear spin $I=3/2$ and is well suited for NMR measurements. A crystal with the dimensions $4.7 \times 1.3 \times 2.5$~mm$^3$ was selected and oriented with its longest [100] axis along the field direction. The experiments were performed at different frequencies; these correspond to different fields for temperatures between 1.5 and 150~K. $^{11\kern-.5pt}$B ($I=3/2$, $\gamma = 8.58406$~MHz/T, 99.88~at.\,\% enrichment) NMR spectra were collected by using a commercial NMR probe over broad field ranges with a \textsc{Tecmag} spectrometer. The spectra were acquired by collecting spin echoes as a function of field, and spin-lattice relaxation measurements were conducted by observing the spin echo following a saturation recovery pulse sequence at the central transition.

Figure~\ref{Fig:NMR}\,(a) shows $^{11}$B NMR spectra collected at several representative temperatures. Note that we have only presented the field sweep spectra collected at the frequency of 33\,MHz. At other frequencies, the spectral behavior is qualitatively similar. Crystallographically, there is only one boron site in~the ludwigite structure of Cu$_2$GaBO$_5$, and therefore in the paramagnetic state one would expect three resonance lines representing different transitions. Indeed, the individual $^{11}$B spectrum consists of one central line, corresponding to the $1/2 \leftrightarrow -1/2$ transition, and two satellites, corresponding to the $1/2 \leftrightarrow 3/2$ and $-1/2 \leftrightarrow -3/2$ transitions. Information about the static internal field distribution and the nature of the magnetic ordering can be obtained from the NMR line width. In the paramagnetic state and at high temperatures, the $^{11}$B field-sweep spectra are quite narrow, and the satellite structure is clearly evident. However, upon lowering the temperature, we observe gradual broadening of all the lines, whereas below 4~K the spectra broaden significantly. This considerable line broadening results from the presence of local static internal field at the $^{11}$B nuclear site corresponding to either short- or long-range static magnetic ordering of the system. This featureless form of the spectrum is an evidence of a broad distribution of internal fields, which may result either from incommensurate magnetic order or from a short-range spin-glass type state with frozen magnetic moments. This interpretation agrees with neutron diffraction, which observes only short-range spin correlations above 1~T.

In addition to the spectra, we have also measured the spin-lattice relaxation rate, $^{11}(1/T_1)$, at the central transition. The NMR spin-lattice relaxation rates probe the dynamical spin susceptibility, providing information about the low-energy excitations and the degree of disorder in the system. The magnetization recovery curves were fit to a stretched exponential function appropriate for the central transition of a spin $I=3/2$ nucleus: $M(t)= M_0\left(1-A\,\mathrm{exp}\bigl[-(\lambda t)^{\beta}\bigr]\right)$, where $M_0$ is the equilibrium nuclear magnetization, and $\beta$ is the stretching exponent. The value of $\beta$ is a measure of disorder of a system. Temperature dependence of the spin-lattice relaxation, $1/{T_1}(T)$, is presented in Fig.~\ref{Fig:NMR}\,(b) at three different values of the external static magnetic field. It can be expressed through the dynamical spin susceptibility by the following relationship~\cite{Moriya63}:
\begin{equation}
\label{eqn:dynamical_susceptibility}
\frac{1}{T_1(T)} = \gamma^2 k_{\rm B}T \lim_{\omega \rightarrow 0} \sum\limits_{\mathbf{q}} \mathcal{F}(\mathbf{q}) \frac{\textrm{Im}\chi(\mathbf{q},\omega,T)}{\hslash \omega},
\end{equation}
where $\mathcal{F}(\mathbf{q})$ is the form factor that depends on the hyperfine coupling tensor, and $\chi(\mathbf{q},\omega,T)$ is the dynamic spin susceptibility.

With decreasing temperature, $1/{T_1}(T)$ at first decreases linearly, but then shows a sharp increase below approximately 8~K. This reflects the slowing down of critical fluctuations near $T_{\rm N}$. The peak related to the AFM transition is sharp at 0.744~T, but gets suppressed and acquires additional broadening with increasing magnetic field. This observation is consistent with the field-induced smearing of the phase transition and destruction of long-range magnetic order observed in our specific-heat and neutron-diffraction data. Note that the increase in $1/{T_1}(T)$ precedes the gradual onset of the diffuse elastic intensity in neutron scattering, which starts at a slightly lower temperature [cf. Fig.~\ref{Fig:E2_peaks}\,(d), where the data were measured in a similar field of 2~T]. Most likely, the spin fluctuations responsible for the enhanced spin-lattice relaxation rate are identical to those observed in the cold-neutron INS data in Fig.~\ref{Fig:NeutronInelastic}.

In the paramagnetic phase, we observe that the spin-lattice relaxation rate increases with the applied magnetic field for $B \lesssim 1$~T. Under the assumption of field-independent dynamic susceptibility, the increase in $1/T_1(T)$ as a function of external field in the paramagnetic state would suggests that $\textrm{Im}\chi$ increases more than linearly with respect to the NMR frequency. Alternatively, it can originate from the strong sensitivity of $\textrm{Im}\chi$ to magnetic fields of the order of 1~T. For comparison, the iron-jarosite AFM kagome system KFe$_3$(OH)$_6$(SO$_4$)$_2$ shows frequency-independent behavior of $1/T_1$ in the paramagnetic state~\cite{NishiyamaMaegawa03}. However, such a frequency dependence has been recently observed in another kagome-lattice system, Fe$_4$Si$_2$Sn$_7$O$_{16}$~\cite{DengreUnpbl}. This difference may be related to the effective Curie-Weiss temperatures $\Theta_\text{CW} \propto J$ in these systems. In the iron jarosite, $\Theta_\text{CW} \approx -800$~K, while for the Fe$_4$Si$_2$Sn$_7$O$_{16}$ system, $\Theta_\text{CW} \approx -12$\,K. The latter is more similar to the ludwigite with its $\Theta_\text{CW} \approx -70$~K and a strongly reduced ordering temperature as a result of frustration. Weaker exchange interactions and strong magnetic frustration in Cu$_2$GaBO$_5$ suggest that the external field of the order of 1~T is sufficient to change the dynamical properties of the spin system.

\vspace{-3pt}\subsection{Muon spin relaxation ($\mu$SR)}\label{Sec:Muons}\vspace{-2pt}

We now proceed to the discussion of the $\mu$SR results. The measurements were performed at the general purpose surface-muon instrument GPS~\cite{AmatoLuetkens17} at the Swiss Muon Source (S$\mu$S) of the Paul Scherrer Institute (PSI, Switzerland). Two samples consisted of several coaligned Cu$_2$Ga$^{11\kern-.5pt}$BO$_5$ and Cu$_2$Al$^{11\kern-.5pt}$BO$_5$ crystals. The initial muon spin polarization was rotated at 45$^\circ$ to the muon beam, so that the time dependence of the muon asymmetry can be measured both on the forward-backward and the top-bottom pairs of positron detectors.

\begin{figure*}
\centerline{\includegraphics[width=0.75\linewidth]{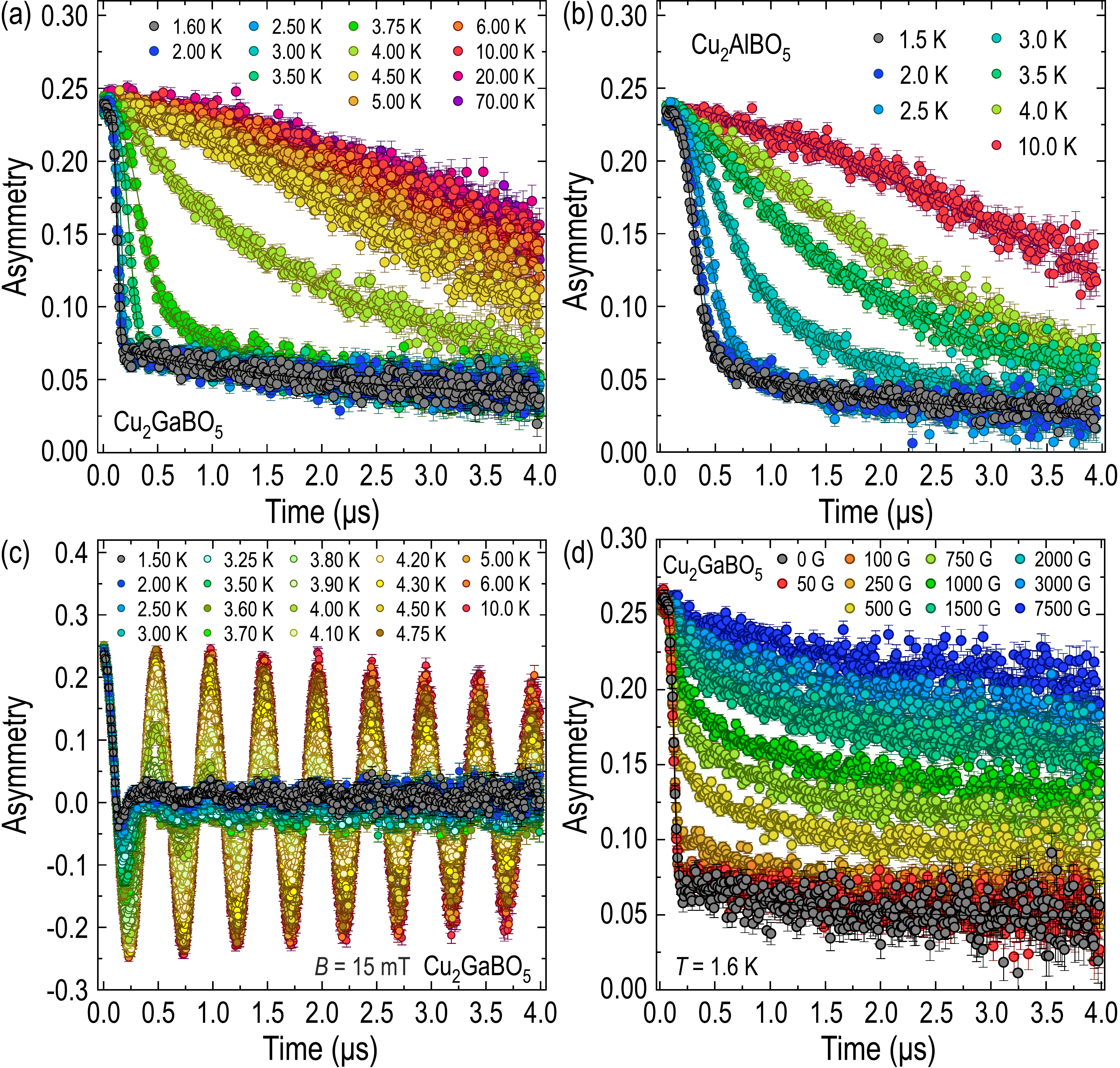}}
\caption{(a,\,b)~Zero-field $\mu$SR time spectra measured at various temperatures for Cu$_2$GaBO$_5$ and Cu$_2$AlBO$_5$. The solid lines are fits to Eq.~(\ref{Eq:Model_Muon_ZF}). (c)~$\mu$SR measurements of the magnetic volume fraction in Cu$_2$GaBO$_5$ as a function of temperature, obtained in a weak transverse magnetic field of 15~mT. The solid lines are fits to Eq.~(\ref{Eq:Model_Muon_TF}). (d)~$\mu$SR time spectra measured in longitudinal fields between zero and 750~mT at the base temperature of 1.6~K. The lines are fits to Eq.~(\ref{Eq:Model_Muon_ZF}).\vspace{-2pt}}
\label{Fig:MuonDatasets}
\end{figure*}

\begin{figure}[b!]\vspace{-7pt}
\centerline{\includegraphics[width=0.8\linewidth]{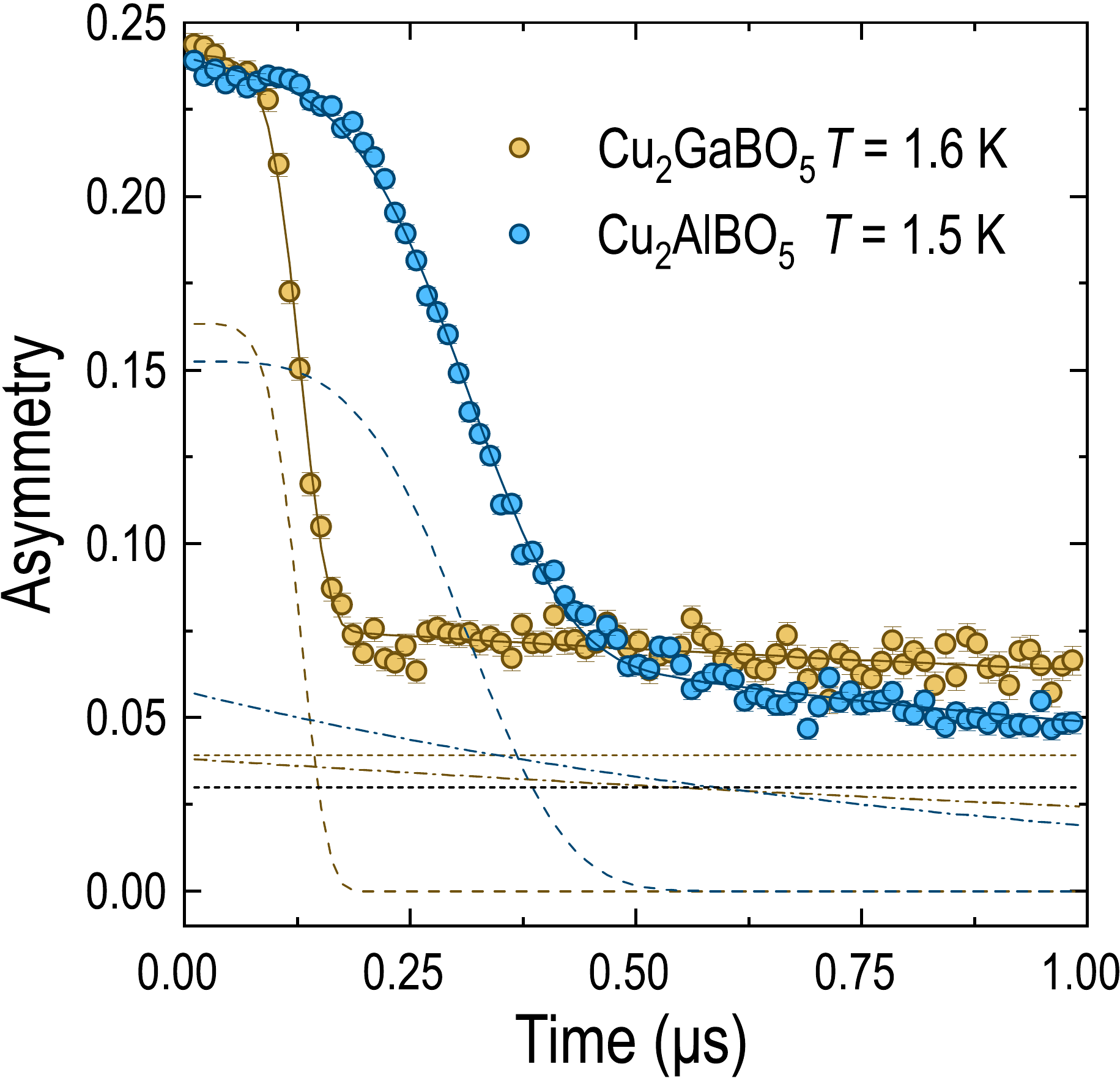}}\vspace{-5pt}
\caption{Comparison of the initial muon depolarization in the time window below 1~$\mu$s at the base temperature for the Cu$_2$GaBO$_5$ and Cu$_2$AlBO$_5$ samples. The solid lines are fits to Eq.~(\ref{Eq:Model_Muon_ZF}), with the individual terms corresponding to the static, dynamic, and background contributions shown with the dashed, dash-dotted, and dotted lines, respectively.\vspace*{-4pt}}
\label{Fig:Det_fit}	
\end{figure}

The zero-field (ZF) relaxation of muon spins at various temperatures is plotted in Figs.~\ref{Fig:MuonDatasets}\,(a) and (b) for the two samples, respectively. In addition, the initial depolarization region within 1~$\mu$s at the base temperature is compared for the two samples in Fig.~\ref{Fig:Det_fit}. First of all, it is evident that in the AFM state, no oscillations of the muon asymmetry are observed in either compound. Instead, we see a monotonic depolarization, starting with a short (0.1--0.2~$\mu$s) time window characterized by a slow relaxation, followed by a much more rapid depolarization. The absence of oscillations is consistent with the incommensurate magnetic order such as a spin spiral, which leads to a broad distribution of internal magnetic fields on the muon stopping sites in different unit cells. However, the delayed depolarization that is preceded by a nearly flat shoulder is quite unusual. It cannot be fitted with either static or dynamic Kubo-Toyabe relaxation function that describes the depolarization in Gaussian fields. We therefore used the following empirical fitting function to describe the $\mu$SR data in zero field:
\begin{equation}
A(t) = A_\text{st}\!\exp[-(\lambda_\text{st}t)^{\beta_\text{st}}] + A_\text{dyn}\exp[-\lambda_\text{dyn}t]+A_\text{bkg},
\label{Eq:Model_Muon_ZF}
\end{equation}
where $A_\text{st}$ and $A_\text{dyn}$ are the muon asymmetries experiencing relaxation in static and dynamic internal fields; $A_\text{bkg}$\,---\,time-independent background asymmetry; $\lambda_\text{st}$ and $\lambda_\text{dyn}$\,---\,depolarization rates for the static and dynamic parts, respectively; and $\beta_\text{st}$ is an empirical exponent used to describe the flat shoulder in the static part of the $\mu$SR spectra, which we assumed constant (temperature-independent) for every sample. For the Cu$_2$GaBO$_5$ and Cu$_2$AlBO$_5$ compounds, the fitted values of $\beta_\text{st}$ are 5.6 and 3.9, respectively.

\begin{figure*}[t!]\vspace{-2pt}
\centerline{\includegraphics[width=\linewidth]{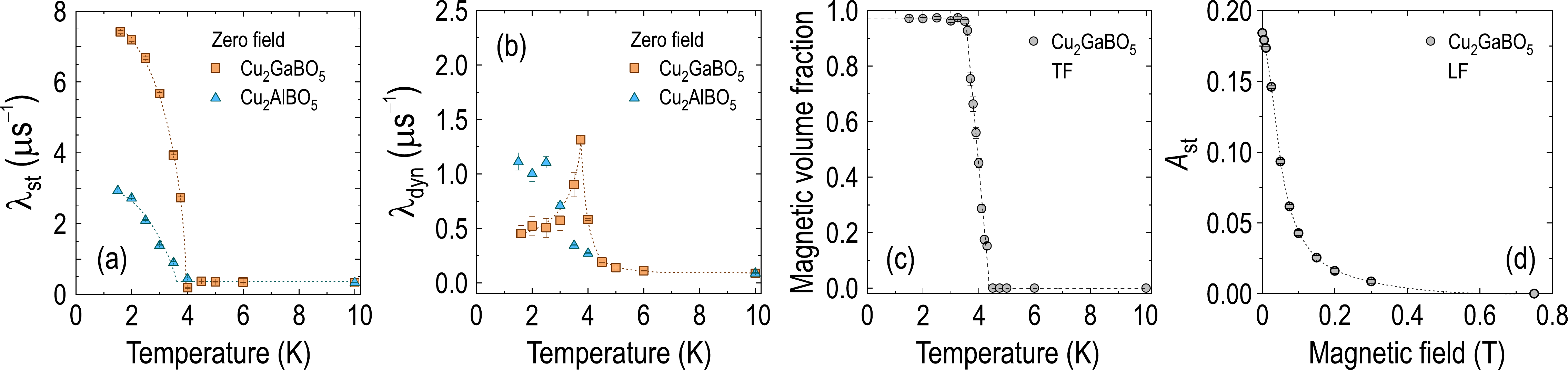}}\vspace{-1pt}
\caption{Temperature dependences of the (a)~static and (b)~dynamic depolarization rates in zero-field $\mu$SR measurements of Cu$_2$GaBO$_5$ and Cu$_2$AlBO$_5$, obtained from fitting Eq.~(\ref{Eq:Model_Muon_ZF}) to the data in Fig.~\ref{Fig:MuonDatasets} as described in the text. (c)~Temperature dependence of the magnetic volume fraction, obtained by fitting the transverse-field data in Fig.~\ref{Fig:MuonDatasets}\,(c) to Eq.~(\ref{Eq:Model_Muon_TF}). (d)~Static part of the muon asymmetry ($A_\text{st}$) as a function of longitudinal magnetic field, obtained by fitting the data in Fig.~\ref{Fig:MuonDatasets}\,(d) to Eq.~(\ref{Eq:Model_Muon_ZF}). The dotted lines are guides to the eyes.\vspace{-1pt}}
\label{Fig:Muon_Param}
\end{figure*}

The depolarization rates $\lambda_\text{st}$ and $\lambda_\text{dyn}$ for both samples, extracted from the fits of ZF datasets in Figs.~\ref{Fig:MuonDatasets}\,(a,\,b), are plotted in Fig.~\ref{Fig:Muon_Param}\,(a,\,b). The static depolarization rate $\lambda_\text{st}$ follows an order-parameter-like temperature dependence in the magnetically ordered state, reaching approximately 3 and 7.5~$\mu$s$^{-1}$ at the lowest measured temperature for the Cu$_2$GaBO$_5$ and Cu$_2$AlBO$_5$ samples, respectively. The longitudinal (dynamic) relaxation rate $\lambda_\text{dyn}$ peaks sharply at $T_\text{N}$ in the Cu$_2$GaBO$_5$ sample, as expected from the critical enhancement of low-energy spin fluctuations near the phase transition, while in the Cu$_2$AlBO$_5$ sample, this anomaly is less pronounced. In both samples, $\lambda_\text{dyn}$ remains high at low temperatures, which may be an indication of persistent low-energy spin fluctuations in the magnetically ordered state. The presence of such gapless spin excitations is also evidenced by our INS data presented in section~\ref{Sec:INS}.

Transverse magnetic field tends to suppress the dynamic part of the muon asymmetry, therefore measurements in a weak transverse field (TF) are sensitive to the magnetic volume fraction in the sample, i.e. the fraction of its volume that has static or slowly fluctuating magnetic moments leading to a rapid depolarization of the muon asymmetry (depolarization rate $\lambda_\text{st}$), as opposed to the paramagnetic volume fraction where the muon asymmetry oscillates slowly with the frequency determined by the external field. Such TF-measurements, shown in Fig.~\ref{Fig:MuonDatasets}\,(c), have been carried out only on the Cu$_2$GaBO$_5$ sample in a small magnetic field $B_{\perp}=15$~mT to assess the width of the AFM phase transition and to estimate the magnetic volume fraction of the sample. The TF-data have been fitted to the following model:
\begin{equation}
\begin{split}
A(t) &= A_0\bigl(\alpha_\text{st}\exp[-(\lambda_\text{st}t)^{\beta_\text{st}}] \\ & + (1-\alpha_\text{st})\exp[-(\lambda_\text{dyn}t)]\bigr)\cos(\gamma_{\!\mu}B_{\perp}t)+A_\text{bkg},
\end{split}
\label{Eq:Model_Muon_TF}
\end{equation}
where $A_0$ is the total asymmetry; $A_\text{bkg}$\,---\,background asymmetry; $\alpha_\text{st}$ and $(1\!-\!\alpha_\text{st})$\,---\,magnetic and paramagnetic volume fractions; $\lambda_\text{st}$ and $\lambda_\text{dyn}$\,---\,depolarization rates for the magnetic and paramagnetic phases, respectively; $\gamma_\mu$\,---\,gyromagnetic ratio of the muon.

Temperature dependence of the magnetic volume fraction $\alpha_\text{st}$, resulting from the fits of the TF data, is plotted in Fig.~\ref{Fig:Muon_Param}\,(c). One can see that the magnetic phase occupies the whole volume of the sample at low temperatures, which excludes any phase separation into magnetically ordered and paramagnetic phases. The width of the magnetic transition is approximately 0.8~K, which may result from a slight smearing of the transition due to the magnetic disorder.

Figure~\ref{Fig:MuonDatasets}\,(d) shows the results of the so-called ``decoupling'' experiment, in which the muon asymmetry on the pair of forward and backward detectors is measured in a weak longitudinal magnetic field. In zero field, the muon depolarization on short time scales results from the random static distribution of internal magnetic fields in the sample, whereas the longitudinal depolarization on longer times scales ($\gtrsim 0.3$~$\mu$s) is determined by dynamical spin fluctuations. The application of a sufficiently high longitudinal field leads to a decoupling of the muon spins from the static internal field distribution, which can be seen as a reduction of the static part of the muon asymmetry, $A_\text{st}$, plotted in Fig.~\ref{Fig:Muon_Param}\,(d). The static part is suppressed already at 0.3~T, which gives us an estimate of the typical width of the internal field distribution on the muon stopping site. On the other hand, slow relaxation due to the dynamic spin fluctuations persists above this field at least up to 0.75~T, as one can see from the nearly unchanged shape of the slowly relaxing part of the $\mu$SR spectra in Fig.~\ref{Fig:MuonDatasets}\,(d). This is consistent with our INS data (Fig.~\ref{Fig:NeutronInelastic}), which show that low-energy quasielastic spin fluctuations persist up to much higher magnetic fields of at least 10~T.

\vspace{-4pt}\section{Density functional theory calculations}\vspace{-2pt}
\label{Sec:Theory}

The observed magnetic propagation vector of Cu$_2$GaBO$_5$ cannot be rationalized within a simplified coupled three-leg ladders model and calls for a microscopic analysis. To this end, we performed density-functional theory (DFT) band-structure calculations within the generalized gradient approximation (GGA) as implemented in the full-potential code FPLO version~18~\cite{KoepernikEschrig99}.

The intersite Cu/Ga disorder in Cu$_2$GaBO$_5$ renders band-structure calculations challenging. To keep the problem tractable, we constructed all realistic ordered configurations based on the crystal structure (Sec.~\ref{sec:str}) and experimental observations. Since our neutron scattering data indicated a nearly antiferromagnetic order along the $a$ axis, we doubled the unit cell along this direction. Based on the X-ray diffraction data, we assumed that $M(1)$ and $M(2)$ positions are fully occupied with Cu. Hence, $M(3)$ and $M(4)$ positions accommodate $\frac13$Cu and $\frac12$Ga. Keeping the Cu\,:\,Ga~=~1\,:\,2 ratio within each position separately is impossible, because the respective multiplicities (4 and 8) are not multiples of 3. Hence, we considered all $12!/4!/(12-4)! = 495$ configurations with the four Cu atoms distributed over $M(3)$ and $M(4)$ positions, preserving the Cu\,:\,Ga stoichiometry within the unit cell.

After symmetrization and removal of duplicates, performed using the \textsc{findsym} program version~6~\cite{oj:findsym01, StokesHatch05}, we were left with 77 unique structures. For all these structures, we calculated the GGA total energies on a 4(8)$\times$4$\times$4 mesh of $k$-points for configurations that do (not) require doubling of the cell along $a$. Nonmagnetic supercell calculations often show poor convergence due to resilient charge redistributions between identical structural blocks. It is the case for Cu$_2$GaBO$_5$: While we were able to converge most configurations, five calculations failed to meet the convergence criteria.

\begin{figure}[t]\vspace{5pt}
\includegraphics[width=\linewidth]{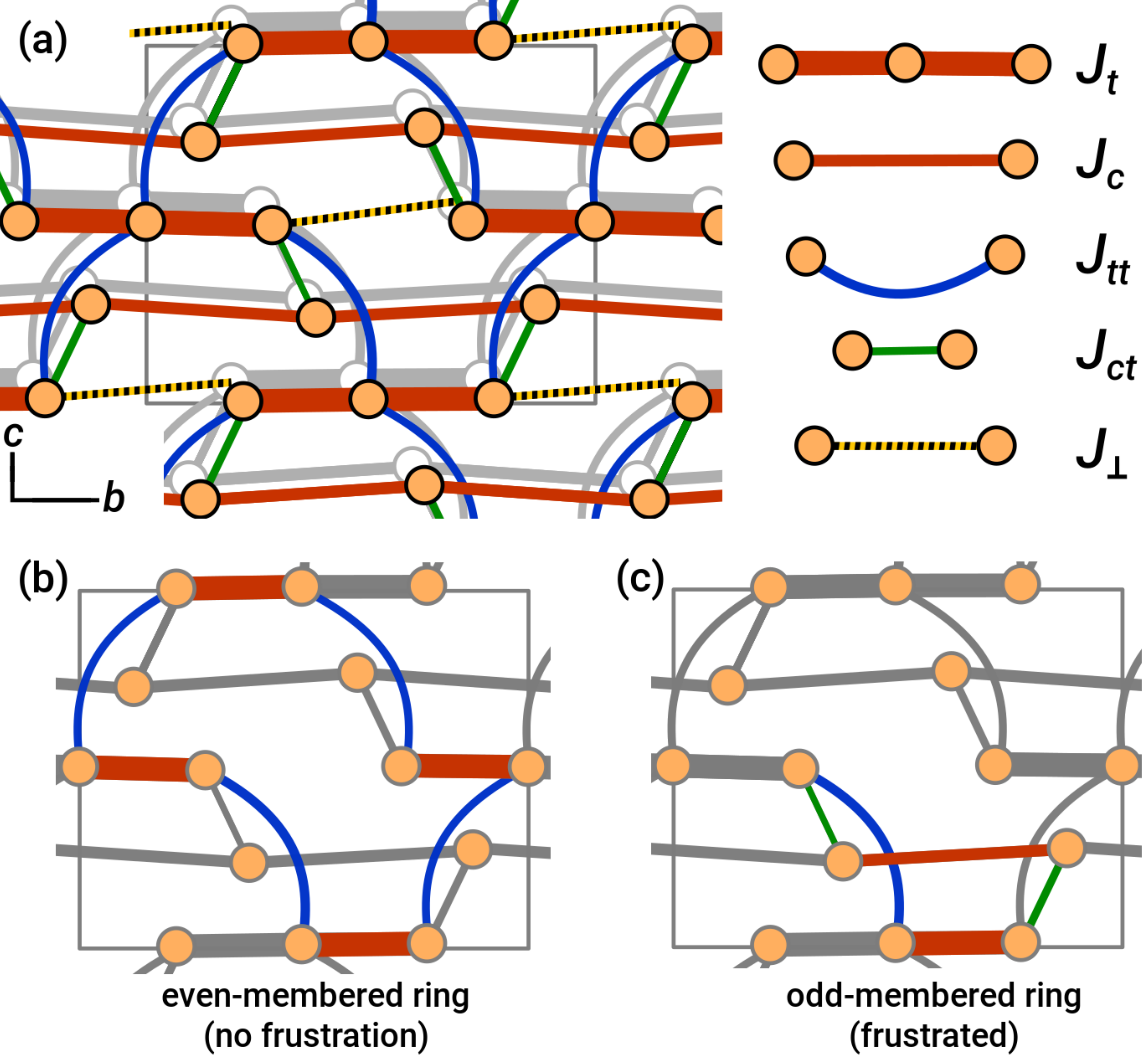}
\caption{(a)~Spin model of Cu$_2$GaBO$_5$ comprising four exchanges\,---\,\jt, \jc, \jtt, and \jct\,---\,that operate within the (102) planes, and the interplane exchange \jp. Spins of the front layer are depicted as orange (shaded) spheres, spins of the second plane are shown with open circles. All exchanges are antiferromagnetic. (b)~Intra-trimer \jt\ and inter-trimer \jtt\ exchanges form eight-membered loops that are not frustrated. (c)~Frustration is induced by the combination of \jt\ and \jtt\ with the intrachain exchange \jc\ and the exchange \jct\ coupling trimers to chains, forming five (odd)-membered rings.}
\label{Fig:Model}
\end{figure}

Looking closer at the crystal structures, we found that many of them feature corner-sharing connections ($d_{\text{Cu..Cu}}=3.867$~\AA) with the Cu--O--Cu bridging angle of 158.8$^{\circ}$. By doing magnetic supercell DFT+$U$ calculations for two such configurations, we found a magnetic exchange of about 700~K in both cases. Such a large magnetic exchange is at odds with the experimental magnetic susceptibility data that do not show any upturn at high temperatures and instead obey the Curie-Weiss law with the Weiss temperature of about 70~K~\cite{EreminaGavrilova20}. Therefore, we excluded all respective configurations from the analysis.

The GGA is known to severely underestimate electronic correlations and in the case of Cu$_2$GaBO$_5$ yields a spurious metallic ground state for all configurations. Hence, we selected four configurations with lowest GGA energies, as well as nonconverged configurations, and performed total-energy GGA+$U$ calculations within the ferromagnetic (FM) state. In this way, we found two lowest-lying configurations, which we refer to as configurations A and B. The atomic coordinates for these lowest-energy ordered configurations are given in the Appendix. While the energies of these two configurations differ by approximately 30~meV/cell, all other configurations lie at least 140~meV higher in energy and therefore can be excluded from the analysis.

For configurations A and B, we performed standard GGA band-structure calculations followed by Wannier projections onto the bands of predominantly Cu~$d_{x^2-y^2}$ orbital character. The constructed effective one-orbital low-energy models allowed us to identify the relevant coupling paths. We restricted ourselves to the hopping integrals, whose absolute value exceeds 50~meV: there are 11~(9) such hoppings in configuration A~(B). The respective exchange integrals were calculated by using the DFT+$U$ supercell approach. We lowered the symmetry by choosing the space group $P1$~($P\overline{1}$) for configuration A~(B), and performed total-energy DFT+$U$ calculations for 14~(16) different collinear spin configurations. We used the fully localized limit for the double counting correction; 8.5 and 1~eV were chosen for the onsite Coulomb repulsion ($U_d$) and the onsite Hund's exchange ($J_d$), respectively. Exchange integrals listed in Table~\ref{tab:exchanges} are solutions to the respective redundant linear problems. Interestingly, only five exchanges exceed 15~K, and this handful is the same in both configurations. The calculated Curie-Weiss temperatures and even the ratios between the leading exchanges are similar in both configurations, therefore in the following discussion we do not distinguish between configurations A and B.

\begin{table}[t!]
\caption{\label{tab:exchanges} Magnetic exchanges, $J$, and the corresponding interatomic distances, $d_{\text{Cu..Cu}}$, for the two ordered configurations of Cu$_2$GaBO$_5$ (see Appendix). The estimates are based on GGA+$U$ total-energy calculations with $U_d = 8.5$~eV and $J_d = 1.0$~eV. The fully localized limit was used for the double-counting correction. The Curie-Weiss temperatures ($\Theta_\text{CW}$, bottom row) were estimated for both configurations, neglecting all further-neighbor interactions beyond our 5-exchange model.\smallskip}
\begin{tabular}{c|c|c|c}
\toprule
Exchange & $d_{\text{Cu..Cu}}$ & \multicolumn{2}{c}{$J$ (K)}\\
         & (\AA) & configuration A & configuration B \\
\midrule
\jt\  & 3.3452 & 112 & 98 \\
\jc\  & 5.9962 & 71  & 68 \\
\jtt\ & 5.9752 & 43  & 37 \\
\jp\  & 5.9096 & 27  & 32 \\
\jct\ & 2.9704 & 21  & 30 \\
\midrule
\multicolumn{2}{c|}{$\Theta_\text{CW}$~(K)} & 46.4 & 49.2 \\
\bottomrule
\end{tabular}
\end{table}

Based on our DFT analysis, we conclude that the spin model of Cu$_2$GaBO$_5$ is quasi-2D and features five relevant exchanges. Despite the presence of edge-sharing CuO$_4$ units, all relevant exchanges are antiferromagnetic. Two strongest exchanges are $J_t$ forming trimers and $J_c$ forming chains along the $b$ axis, as shown in Fig.~\ref{Fig:Model}\,(a). Trimers are connected to chains by $J_\text{ct}$ and to each other by $J_\text{tt}$. In addition to these four in-plane exchanges, $J_{\perp}$ couples trimers of the neighboring planes [Fig.~\ref{Fig:Model}\,(a)]. The spin model is frustrated, but in an intricate way. Note that intra- and intertrimer exchanges, $J_t$ and $J_{tt}$, do not suffice: their combination gives rise to even-membered rings, and hence a bipartite lattice [Fig.~\ref{Fig:Model}~(b)]. Only if we include all four in-plane exchanges, odd-membered rings emerge in the spin lattice, such as the five-membered ring in Fig.~\ref{Fig:Model}\,(c).

Magnetic frustration present in the spin model of Cu$_2$GaBO$_5$, which is illustrated in Fig.~\ref{Fig:Model}\,(a), may stabilize different ordered states, collinear or noncollinear, as well as a gapped state. The choice generally depends on topology of exchange paths and the strength of frustration, but in a nontrivial way, even for seemingly simple topologies. Obviously, further insights into the magnetic ground state and the excitation spectrum of this model are highly desirable. However, they require large-scale exact diagonalization or density matrix renormalization group (DMRG) studies, beyond the scope of this paper.

\vspace{-2pt}\section{Discussion and conclusions}\vspace{-2pt}

In this work, we have investigated the magnetic properties of copper ludwigites using magnetic neutron diffraction, inelastic neutron scattering, local-probe spectroscopies (NMR, $\mu$SR), and DFT calculations. The most remarkable feature of these compounds is that in spite of strong site disorder, they show long-range magnetic order at zero magnetic field, which can be destroyed already in relatively weak fields with the formation of a spin-glass-like ground state. The effect of magnetic disorder can be therefore continuously tuned by an easily accessible external parameter. While most of our data were presented only for the Cu$_2$GaBO$_5$ compound, the qualitative similarity between our thermodynamic measurements and $\mu$SR results on Cu$_2$GaBO$_5$ and Cu$_2$AlBO$_5$ suggests that the findings must be generic for the whole structural family of homomagnetic copper ludwigites (i.e. those with a nonmagnetic $M^\prime$ ion).

In a conventional antiferromagnet, the application of magnetic field is expected to destabilize the AFM order, which should ultimately lead to a fully field-polarized collinear phase. In frustrated spin system, this often occurs through a sequence of field-driven metamagnetic phase transitions that separate various intermediate field-induced magnetically ordered states. The role of frustration is to weaken the AFM state and bring several competing ordered states closer in energy, so that field-induced quantum phase transitions are observable at relatively low (in comparison to the dominant exchange constant, $J$), experimentally accessible magnetic fields. A prominent example of such behavior is given by the magnetization plateaus in triangular or kagome antiferromagnets~\cite{TeradaNarumi07, OkamotoTokunaga11, IshikawaYoshida15}. Contrary to this expected behavior, the field-driven metamagnetic quantum phase transition is avoided in ludwigites through a crossover to a spin-glass-like state resulting from the rapid destruction of long-range magnetic order in an applied magnetic field.

We speculate that this may happen as a result of competition between two magnetic subsystems residing on the structurally ordered ($M(1)$ and $M(2)$ sites) and structurally disordered ($M(3)$ and $M(4)$ sites) sublattices. It is plausible that the disordered and magnetically dilute sublattices either do not participate in the AFM order, remaining in a paramagnetic state below $T_\text{N}$, or simply inherit the AFM correlations of the ordered sublattice due to the negligibly small number of exchanges between orphan spins. As long as such a disordered subsystem carries no overall magnetization, it has no significant contribution to the overall magnetic energy of the system on average, allowing for the formation of a long-range order on the $M(1)$ and $M(2)$ sublattices. However, as soon as the disordered spins get polarized by an external field, their influence on the ordered subsystem increases sufficiently to suppress the long-range order.

While this qualitative explanation appears plausible, so far we have no direct evidence for the number of Cu$^{2+}$ spins participating in the AFM ground state. It also remains unclear, what minimal model is required to capture the effective coupling between the two magnetic subsystems and its dependence on the external magnetic field. Obviously, the effects of disorder cannot be captured by the DFT calculations presented in section~\ref{Sec:Theory}, as they are restricted to hypothetical ordered crystal configurations. On the other hand, these calculations allowed us to classify copper ludwigites as quasi-2D antiferromagnets with a complex network of magnetic interactions and to estimate the relevant exchange constants, which can be helpful in developing more accurate spin models and in guiding future experiments on this structural family of compounds.

\vspace{-2pt}\section*{Acknowledgments}\vspace{-2pt}

We thank U.~Nitzsche for technical assistance. This project was funded in part by the German Research Foundation (DFG) under Grant No.~IN~209/9-1, via the project C03 of the Collaborative Research Center SFB 1143 (project-id 247310070) at the TU Dresden and the W\"urzburg-Dresden Cluster of Excellence on Complexity and Topology in Quantum Matter\,---\,\textit{ct.qmat} (EXC 2147, project-id 390858490). O.\,J. was supported by the Leibniz Association through the Leibniz Competition.

\onecolumngrid\clearpage

\vspace{-3pt}\section*{Appendix}\vspace{-8pt}

\begin{table*}[h]\vspace{-1em}
\caption{\label{tab:configA} Atomic coordinates of the hypothetical ordered configuration A discussed in section~\ref{Sec:Theory}.\smallskip}
\textbf{Configuration~A:} Space group $P1$~(\#1), $a = 6.2252$~\r{A}, $b = 9.4792$~\r{A}, $c = 11.9215$~\r{A}, $\alpha = \beta = 90^\circ$, $\gamma = 97.909^\circ$.\\\smallskip
\begin{tabular}{lrrr@{\quad}|@{\quad}lrrr@{\quad}|@{\quad}lrrr}
\toprule
site & $x/a$~~~ & $y/b$~~~ & $z/c$~~~ & site & $x/a$~~~ & $y/b$~~~ & $z/c$~~~ & site & $x/a$~~~ & $y/b$~~~ & $z/c$~~~ \\ \midrule
Cu(1) & 0.26764 &  0.49286 &  0.71963 & O(1) & 0.04380 &  0.33270 &  0.76220 & O(25) & 0.27040 &  0.60080 &  0.85600 \\
Cu(2) & 0.76764 &  0.49286 &  0.71963 & O(2) & 0.54380 &  0.33270 &  0.76220 & O(26) & 0.77040 &  0.60080 &  0.85600 \\
Cu(3) & 0.23236 &  0.50714 &  0.28037 & O(3) & 0.45620 &  0.66730 &  0.23780 & O(27) & 0.22960 &  0.39920 &  0.14400 \\
Cu(4) & 0.73237 &  0.50714 &  0.28037 & O(4) & $-0.04380$ & 0.66730 &  0.23780 & O(28) & 0.72960 &  0.39920 &  0.14400 \\
Cu(5) & 0.23237 &  0.00714 &  0.21963 & O(5) & 0.45620 &  0.16730 &  0.26220 & O(29) & 0.22960 &  0.89920 &  0.35600 \\
Cu(6) & 0.73237 &  0.00714 &  0.21963 & O(6) & $-0.04380$ & 0.16730 &  0.26220 & O(30) & 0.72960 &  0.89920 &  0.35600 \\
Cu(7) & 0.26764 &  $-0.00714$ & 0.78037 & O(7) & 0.04380 &  0.83270 &  0.73780 & O(31) & 0.27040 &  0.10080 &  0.64400 \\
Cu(8) & 0.76763 &  $-0.00714$ & 0.78037 & O(8) & 0.54380 &  0.83270 &  0.73780 & O(32) & 0.77040 &  0.10080 &  0.64400 \\
Cu(9) & 0.25000 &  0.50000 &  0.00000 & O(9) & 0.48595 &  0.15720 &  0.46170 & O(33) & 0.24620 &  0.38970 &  0.57730 \\
Cu(10) & 0.75000 &  0.50000 &  0.00000 & O(10) & $-0.01405$ & 0.15720 &  0.46170 & O(34) & 0.74620 &  0.38970 &  0.57730 \\
Cu(11) & 0.25000 &  0.00000 &  0.50000 & O(11) & 0.01405 &  0.84280 &  0.53830 & O(35) & 0.25380 &  0.61030 &  0.42270 \\
Cu(12) & 0.75000 &  0.00000 &  0.50000 & O(12) & 0.51405 &  0.84280 &  0.53830 & O(36) & 0.75380 &  0.61030 &  0.42270 \\
Cu(13) & 0.03599 &  0.22804 &  0.61903 & O(13) & 0.01405 &  0.34280 &  $-0.03830$ & O(37) & 0.25380 &  0.11030 &  0.07730 \\
Cu(14) & 0.46401 &  0.77196 &  0.38097 & O(14) & 0.51405 &  0.34280 &  $-0.03830$ & O(38) & 0.75380 &  0.11030 &  0.07730 \\
Cu(15) & 0.46400 &  0.27196 &  0.11903 & O(15) & 0.48595 &  0.65720 &  0.03830 & O(39) & 0.24620 &  0.88970 &  $-0.07730$ \\
Cu(16) & 0.53599 &  0.72804 &  0.88097 & O(16) & $-0.01405$ & 0.65720 &  0.03830 & O(40) & 0.74620 &  0.88970 &  $-0.07730$ \\
Ga(1) & 0.53599 &  0.22804 &  0.61903 & O(17) & 0.49720 &  0.61900 &  0.63420 & B(1) & 0.48240 &  0.23480 &  0.36460 \\
Ga(2) & $-0.03600$ & 0.77196 &  0.38097 & O(18) & $-0.00280$ & 0.61900 &  0.63420 & B(2) & $-0.01760$ & 0.23480 &  0.36460 \\
Ga(3) & $-0.03599$ & 0.27196 &  0.11903 & O(19) & 0.00280 &  0.38100 &  0.36580 & B(3) & 0.01760 &  0.76520 &  0.63540 \\
Ga(4) & 0.03599 &  0.72804 &  0.88097 & O(20) & 0.50280 &  0.38100 &  0.36580 & B(4) & 0.51760 &  0.76520 &  0.63540 \\
Ga(5) & 0.00000 &  0.50000 &  0.50000 & O(21) & 0.00280 &  0.88100 &  0.13420 & B(5) & 0.01760 &  0.26520 &  0.86460 \\
Ga(6) & 0.50000 &  0.50000 &  0.50000 & O(22) & 0.50280 &  0.88100 &  0.13420 & B(6) & 0.51760 &  0.26520 &  0.86460 \\
Ga(7) & 0.00000 &  0.00000 &  0.00000 & O(23) & 0.49720 &  0.11900 &  0.86580 & B(7) & 0.48240 &  0.73480 &  0.13540 \\
Ga(8) & 0.50000 &  0.00000 &  0.00000 & O(24) & $-0.00280$ & 0.11900 &  0.86580 & B(8) & $-0.01760$ & 0.73480 &  0.13540 \\
\bottomrule
\end{tabular}
\end{table*}

\begin{table*}[h]\vspace{-1em}
\caption{\label{tab:configB} Wyckoff positions and atomic coordinates of the hypothetical ordered configuration B discussed in section~\ref{Sec:Theory}.\smallskip}
\textbf{Configuration~B:} Space group $P2_1/c$~(\#14); $a = 6.2252$~\r{A}, $b = 11.9215$~\r{A}, $c = 10.6004$~\r{A}; $\beta = 117.659^\circ$.\\\smallskip
\begin{tabular}{lc@{}rrr@{\quad}|@{\quad}lc@{}rrr}
\toprule
site & Wyckoff position & $x/a$~~~ & $y/b$~~~ & $z/c$~~~ & site & Wyckoff position & $x/a$~~~ & $y/b$~~~ & $z/c$~~~\\ \midrule
Cu(1) & $4e$ & 0.02477 &  0.21963 &  0.00714 & O(1) & $4e$ & $-0.03890$ & 0.26220 &  0.16730 \\
Cu(2) & $4e$ & 0.52477 &  0.21963 &  0.00714 & O(2) & $4e$ & 0.46110 &  0.26220 &  0.16730 \\
Cu(3) & $2c$ & 0.00000 &  0.00000 &  0.50000 & O(3) & $4e$ & 0.57875 &  $-0.03830$ & 0.34280 \\
Cu(4) & $2d$ & 0.50000 &  0.00000 &  0.50000 & O(4) & $4e$ & 0.07875 &  $-0.03830$ & 0.34280 \\
Cu(5) & $4e$ & 0.05795 &  0.11903 &  0.27196 & O(5) & $4e$ & 0.12820 &  0.13420 &  0.88100 \\
Ga(1) & $4e$ & 0.55795 &  0.11903 &  0.27196 & O(6) & $4e$ & 0.62820 &  0.13420 &  0.88100 \\
Ga(2) & $4e$ & 0.75000 &  0.00000 &  0.00000 & O(7) & $4e$ & $-0.08040$ & 0.35600 &  0.89920 \\
B(1) & $4e$ & 0.49760 &  0.86460 &  0.26520 & O(8) & $4e$ & 0.41960 &  0.35600 &  0.89920 \\
B(2) & $4e$ & $-0.00240$ & 0.86460 &  0.26520 & O(9) & $4e$ & 0.10650 &  0.07730 &  0.11030 \\
     &      &            &         &          & O(10) & $4e$ & 0.60650 &  0.07730 &  0.11030 \\
\bottomrule
\end{tabular}\vspace{-1em}
\end{table*}

\end{document}